# A dynamic ridesharing dispatch and idle vehicle repositioning strategy with integrated transit transfers

Tai-Yu Ma[a], Saeid Rasulkhani[b], Joseph Y. J. Chow[b], Sylvain Klein[a]

[a]LISER, 11 Porte des Sciences, L-4366 Esch-sur-Alzette, Luxembourg
[b]C²SMART University Transportation Center, New York University, 6 MetroTech Center, RH 400B Brooklyn, NY 11201, USA

**Abstract**

We propose a ridesharing strategy with integrated transit in which a private on-demand mobility service operator may drop off a passenger directly door-to-door, commit to dropping them at a transit station or picking up from a transit station, or to both pickup and drop off at two different stations with different vehicles. We study the effectiveness of online solution algorithms for this proposed strategy. Queueing-theoretic vehicle dispatch and idle vehicle relocation algorithms are customized for the problem. Several experiments are conducted first with a synthetic instance to design and test the effectiveness of this integrated solution method, the influence of different model parameters, and measure the benefit of such cooperation. Results suggest that rideshare vehicle travel time can drop by 40–60% consistently while passenger journey times can be reduced by 50–60% when demand is high. A case study of Long Island commuters to New York City (NYC) suggests having the proposed operating strategy can substantially cut user journey times and operating costs by up to 54% and 60% each for a range of 10–30 taxis initiated per zone. This result shows that there are settings where such service is highly warranted.

# 1. Introduction

There is huge potential for collaborations between public transport agencies and private transport operators to leverage mobility-on-demand (MoD) services (Murphy and Feigon, 2016). The basic form of collaboration is for MoD services to cover the first and last mile segments of a passenger trip. This is becoming increasingly popular, as shown in Table 1 (e.g. Quadrifoglio and Li, 2009; Wang and Odoni, 2016; Djavadian and Chow, 2017; Guo et al., 2017; Shen et al., 2018). These initiatives suggest such partnerships can provide better connectivity and improve the efficiency and flexibility of the coexisting fixed-route transit service.

Table 1. Examples of public-private partnerships with mobility services to address last mile problem

| Public agency | Private company | Project | Source |
| --- | --- | --- | --- |
| Helsinki | Kutsuplus | On-demand minibus | (Wired, 2013) |
| Dallas Area Rapid Transit | Lyft | Dallas | (DART, 2015) |
| JFK Airport | Bandwagon | Cab carpool | (Daily News, 2015) |
| Kansas City | Bridj | Microtransit service | (Kansas City Star, 2015) |
| Los Angeles Airport | Lyft | LAX access | (The Verge, 2015) |
| Metrolinx | RideCo | Last mile | (CBC, 2015) |
| Amtrak | Lyft | Last mile | (TechCrunch, 2017) |
| Arlington, TX | Via | On-demand minibus | (TechCrunch, 2018) |
| San Francisco | Chariot | Private transit | (SF Chronicle, 2018) |

However, the basic structure does not coordinate the multimodal segments of a passenger's trip. There is no *integrated* optimization of vehicle dispatch and repositioning of idle vehicles with transit stations to provide an integrated, multimodal trip. This second, more sophisticated, collaborative structure between the MoD operator and transit agency is shown in Fig. 1. The presence of public transport creates a broader array of options for using rideshare: rideshare from door-to-door (R), rideshare-to-transit-to-rideshare (RTR), rideshare-to-transit-to-walk (RTW) or vice versa (WTR). The passenger gets a seamless service option in which a single fare is paid, likely at a much more discounted rate than if they were dropped off door-to-door by the operator (especially if the distance is far enough and well-served by an existing transit system). The transit system gets higher ridership and can serve riders that may typically be discouraged by the high last mile access costs. Lastly, the operator saves on operating costs for transporting along a path that is already well served by existing transit system capacity.

Some integrated trip planning tools exist to provide multimodal trip information to passengers (e.g. TriMet in Portland, and only for certain modes like biking and transit, TriMet News, 2012) but there is no integrated dispatch and fleet management algorithm. We hypothesize that the benefits of integrated service are highly dependent on dynamic operations and are not trivial to integrate. But how much would such algorithms benefit such an operational strategy? What algorithmic designs need to be proposed to make the integrated system work?

Two primary contributions are made to address those questions. First, we propose and design a new rideshare service strategy that provides end-to-end service while leveraging transfers to/from coexisting transit networks. Second, we modify and test different integration designs of state-of-the-art anticipatory dispatch and relocation algorithms to work together in this system, resulting in a number of new proposed features. Computational experiments are conducted in synthetic instances as well as in a large-scale case study of the Long Island Railroad (LIRR) accessing New York City (NYC) to provide insights on how to select algorithm parameters to obtain effective results.

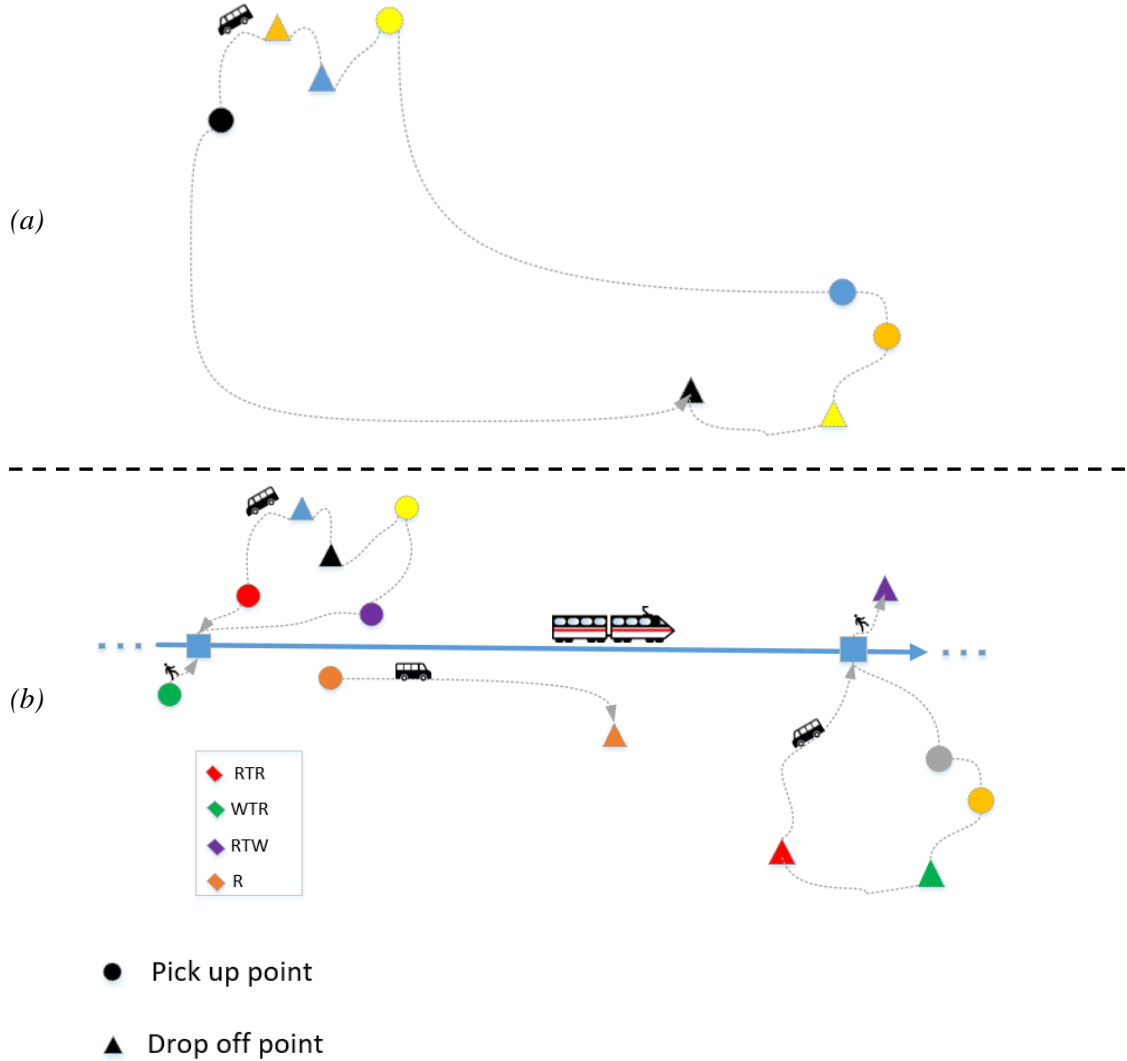

Fig. 1. Illustration of classical DARP (a) and bimodal ridesharing in collaboration with a coexisting transit system (b).

## 2. Literature Review

    Different studies have been conducted on private service operations in the presence of public transit systems. Chow and Sayarshad (2014) proposed a symbiotic framework for network design problems in proximity to different transportations networks. Mobility-as-a-Service (MaaS) (Djavadian and Chow, 2017; Hensher, 2017; Catapult Transport Systems, 2016) is an emerging paradigm where multiple mobility operators and technology providers work together to serve a trip, even if it may involve multiple modes made on-demand. Martinez and Viegas (2017) studied the impact of using a shared self-driving system using an agent-based method. The simulation is done in the presence of a metro system. The study is a demand evaluation of such a strategy, not a comparison of algorithmic design study.

    In dynamic MoD systems, decisions are made over time and demand is not known in advance. Most studies of this type focus on uni-modal vehicle dispatching and routing policy design on road networks (Furuhata et al., 2013; Sayarshad and Chow, 2015). Different exact and approximation methods have been proposed for solving static and dynamic dial-a-ride problems (DARP) (Braekers et al., 2014; Jaw et al., 1986; Parragh et al., 2008; Kirchler and Wolfler Calvo, 2013; Liu et al., 2014). Among dynamic routing and dispatch, some studies have considered non-myopic or anticipative

strategies (e.g. Bent and Van Hentenryck, 2004; Thomas, 2007; Ichoua et al., 2006; Hyytiä et al., 2012; Sayarshad and Chow, 2015). These studies generally consider a system in isolation from other operators.

To anticipate future states of the system and make optimal routing decisions, a Markov decision process provides a theoretical framework to model DARP policies under a stochastic setting (Howard, 2007). Determining the expected value requires full specification of future states of the system, which becomes an intractable problem. Approximate dynamic programming (ADP) methods have been proposed (Secomandi, 2001; Ulmer, 2017). However, these approximation methods tend to be limited to one or two step look-ahead (see Sayarshad and Chow, 2015). Hyytiä et al. (2012) proposed an infinite horizon approximation of the expected value of future states of the system to solve the DARP. They showed the dispatching and routing policy can effectively reduce overall operating cost and customers' riding time (Hyytiä et al., 2012; Sayarshad and Chow, 2015), although poor performances can also be observed in some cases (Chow and Sayarshad, 2016). The non-myopic vehicle dispatching policy is to assign a vehicle $v^{opt}$ with updated post-decision route $x_t^{v*}$ based on minimizing the additional insertion cost of a new request among all practically nearby vehicles, considering an approximation of the future cost as a M/M/1 queue delay (Hyytiä et al., 2012) in Eq. (1).

$$\{v^*, x_t^{v*}\} = \mathrm{argmin}_{v,x}[c(v, x_t^v) - c(v, x)] \qquad (1)$$

where $x_t^v$ is a new tour after inserting a new request, and $c(v, x)$ is a cost function for vehicle $v$ operating tour $x$, shown in Eq. (2).

$$c(v, x) = \gamma T(v, x) + (1 - \gamma)[\beta\, T(v, x)^2 + \sum_{n \in P_v} Y_n(v, x)] \qquad (2)$$

where $T(v, x)$ is the length (measured in time) of tour $x$ and $Y_n(v, x)$ is the journey time (waiting time plus in-vehicle travel time) for passenger $n$ among the set of passengers $P_v$ assigned to vehicle $v$. $T(v, x)$ is related to system cost. $\sum_{n \in P_v} Y_n(v, x)$ is related to customers' inconvenience. Hyytiä et al. (2012) derived the expression for delay for the dispatch problem when operating as an M/M/1 queue and found it proportional to $T(v, x)^2$ (i.e. $\frac{\mu\lambda}{2(\mu-\lambda)} T(v, x)^2$, to be precise) which is then parameterized as $\beta\, T(v, x)^2$. This queue delay is used to approximate the future, steady state cost (i.e. non-myopic costs). The parameter $\gamma$ is a conversion coefficient between customer cost and system cost while $\beta$ is the degree of look-ahead parameter: when $\beta = 0$, the methodology becomes purely myopic (Hyytiä et al., 2012). The tour length function $T(v, x)$ and tour state $x$ are obtained by solving a traveling salesman problem with pickups and drop-offs (TSPPD). One can apply state-of-the-art heuristics for that portion (Agatz et al., 2012; Parragh et al., 2010; Parragh and Schmid, 2013). Numerical studies show that using the non-myopic vehicle dispatching approach can effectively reduce total system operation cost and average customers' riding time compared to myopic dynamic dispatch and routing (Hyytiä et al., 2012; Sayarshad and Chow, 2015).

Another important issue is related to the idle vehicle relocation problem as it presents a considerable running cost for shared mobility systems (Sayarshad and Chow, 2017; Vogel, 2016). This issue has drawn increasing attention in recent years for shared mobility systems (Bruglieri et al., 2017; Martinez et al. (2015); Nourinejad et al., 2015; Powell et al., 2011; Santos and Correia, 2015; Sayarshad and Chow, 2017; Weikl and Bogenberger, 2015; Boyaci et al., 2017; Jorge et al., 2014; Kek et al., 2009). The idle vehicle relocation problem can also be divided into myopic and non-myopic methods. Yuan et al. (2011) used taxi trajectory data to design a recommendation system for taxi drivers and customers to reduce searching/waiting time of each driver/rider. For a non-myopic idle vehicle relocation policy, Sayarshad and Chow (2017) proposed a queueing-theoretic approach for real-time optimal idle vehicle relocation. Other studies suggest using the queueing-theoretic model to rebalance idle vehicles (Zhang and Pavone, 2016; Spieser et al. 2016) or modeling idle vehicle rebalancing with continuous approximation (Pavone et al., 2012; Li et al., 2016). Like with the dynamic routing literature, the non-myopic studies, as highlighted in this review, tend to focus on single operator systems absent of other operators.

A subset of the static and dynamic MoD literature deals with last mile access to transit services. Liaw et al. (1996) considered a bimodal dial-a-ride problem and proposed a linear mixed integer programming model to find optimal vehicle routes and schedules for paratransit service. Cangialosi et al. (2016) proposed a mixed integer linear programming model for multimodal trips with intermodal transfers. Ghilas et al. (2016) considered an integrated freight transport service using pick-up and delivery vehicles and fixed-route public transport to design least cost routes. Masson et al. (2014) considered a static dial-a-ride problem with the presence of a set of transfer points and proposed an adaptive large neighborhood search metaheuristic. Stiglic et al. (2018) considered rideshare-transit cooperation by optimizing rideshare as a last mile option to drop passengers at a transit station. Some recent studies for addressing last-mile passenger transportation problem with integrated public transport system can be found in Raghunathan et al. (2018), Stiglic et al. (2018), and Wang (2017). However, these studies only consider the access problem and not door-to-door multimodal optimization.

The non-myopic optimal idle vehicle relocation model of Sayarshad and Chow (2017) is recalled as follows. The problem is considered as a multiple server location problem under stochastic demand. We rebalance the locations of idle vehicles given stochastic demand such that total rebalancing operation cost, customers' inconvenience (travel time), and an infinite horizon future cost of serving customers (modeled as a queue delay) are minimized. Let the entire service region $\mathcal{A}$ be divided into a set of zones $\bar{N}$. The idle vehicle rebalancing is executed at the beginning of each relocation time interval (epoch), i.e. a couple of minutes. The objective is to assign idle vehicles between zones at each relocation epoch $h$ under a set of vehicle flow conservation constraints as follows.

Table 2 Notation for the idle vehicle relocation problem

| | |
|---|---|
| $\bar{N}$ | Set of zones |
| $\lambda_i$ | Arrival rate at zone $i$ during last relocation epoch $h-1$ |
| $\mu_j$ | Service rate at zone $j$ during last relocation epoch $h-1$ |
| $t_{ij}$ | Travel time from zone $i$ to zone $j$ |
| $r_{ij}$ | Relocation cost from zone $i$ to zone $j$ |
| $B$ | Number of total idle vehicles at the beginning of epoch $h$ (index h is dropped) |
| $C_j$ | Maximum possible number of idle vehicles at zone $j$ |
| $y_j$ | Number of idle vehicles at zone $j$ at the beginning of relocation epoch $h$ (index $h$ is dropped) |
| $\theta$ | A conversion scalar |
| $\rho_{\eta m}$ | Utilization rate constraints for a reliability level $\eta$ for having $m$ servers |
| *Decision variable* | |
| $W_{ij}$ | Flow of idle vehicle relocation from zone $i$ to zone $j$ for relocation epoch $h$ (index $h$ is dropped) |
| $X_{ij}$ | Customers arrive at zone $i$ served by idle vehicle at zone $j$ if set as 1 |
| $Y_{jm}$ | $m$-th idle vehicle located at zone $j$ comes to serve customers if set as 1 |
| $S_i$ | Dummy variable representing the supply of idle vehicles from zone $i$ |
| $D_j$ | Dummy variable representing the demand of idle vehicles to zone $j$ |

$$\Phi = min \sum_{i \in \bar{N}} \sum_{j \in \bar{N}} \lambda_i t_{ij} X_{ij} + \theta \sum_{i \in \bar{N}} \sum_{j \in \bar{N}} r_{ij} W_{ij} \tag{3}$$

Subject to:
$$\sum_{j \in \bar{N}} X_{ij} = 1, \quad \forall i \in \bar{N} \tag{4}$$

$$Y_{jm} \leq Y_{j,m-1}, \quad \forall j \in \bar{N}, m = 2,3,\dots,C_j \tag{5}$$

$$\sum_{i\in\bar{N}} \lambda_i X_{ij} \le \mu_j \left[ Y_{j1}\rho_{\eta j1} + \sum_{m=2}^{C_j} Y_{jm}(\rho_{\eta jm} - \rho_{\eta j,m-1}) \right], \quad \forall j \in \bar{N} \tag{6}$$

$$X_{ij} \le Y_{j1}, \quad \forall i,j \in \bar{N} \tag{7}$$

$$\sum_{j\in\bar{N}} \sum_{m=1}^{C_j} Y_{jm} = B \tag{8}$$

$$\sum_{j\in\bar{N}} W_{ij} = S_i, \quad \forall i \in \bar{N} \tag{9}$$

$$\sum_{i\in\bar{N}} W_{ij} = D_j, \quad \forall j \in \bar{N} \tag{10}$$

$$S_j \le y_j, \quad \forall j \in \bar{N} \tag{11}$$

$$-D_j + S_j - y_j + \sum_{m=1}^{C_j} Y_{jm} \le 0, \quad \forall j \in \bar{N} \tag{12}$$

$$X_{ij} \in \{0,1\} \tag{13}$$

$$0 \le Y_{jm} \le 1 \tag{14}$$

$$D_j, S_j \ge 0, W_{ij} \in Z^+ \tag{15}$$

The objective function minimizes customer access time to idle vehicles and total idle vehicle relocation cost. Eq. (4) requires that customers at zone $i$ be served by idle vehicles from zone $j$. Note that this assumes, consistent with the model formulation in the literature, that all customers in zone $i$ are assigned to be served by one zone $j$. If we were to alternatively allow users in zone $i$ be served by multiple different zones, it would require changing the location model into a much more complex location problem with equilibrium constraints.

Eq. (5) is an order constraint which states the (m-1)-th idle vehicle is relocated before m-th idle vehicle. Eq. (6) is a queue intensity constraint that requires no more than b other customers waiting on a line with a probability more than service reliability $\eta$. The higher the value of $\eta$, the lower the queue delay allowed for customers. Eq. (7) ensures the allocation of customers to only an idle vehicle. Eq. (8) ensures total available idle vehicles. Eq. (9) and (10) represent the supply and demand of idle vehicle flows. Eq. (11) ensures initial available idle vehicles at node j must equal or be greater than total relocated idle vehicles from $j$. Eq. (12) ensures total idle vehicles at node $j$ after relocation must be equal or greater than total vehicles from $j$ to serve customers. Eq. (11) and Eq. (12) are new formulations replacing two flow conservation equations in Sayarshad and Chow (2017). This new formulation considers both inflow and outflow of idle vehicles relocating to and from a node to model flow conservation by considering total flow movements at that node. Eq. (13-15) are binary and non-negativity constraints. The intensity parameter $\rho_{\eta jm}$ in Eq. (6) is determined exogenously by finding the root of Eq. (16) (Marianov and Serra, 1998, 2002; Sayarshad and Chow, 2017) for each combination of values $\eta$, m, and b ($\eta$ and b are user-defined parameters) and input in the model as a parameter. The reader is referred to Marianov and Serra (2002) for more detail.

$$\sum_{k=0}^{m-1} \left( (m-k)m!\, m^b/k! \right) \left( 1/\rho^{m+b+1-k} \right) \ge 1/(1-\eta) \tag{16}$$

The queue delay is used as an approximation of future costs in a non-myopic context. If we relax Eq. (6) the model becomes myopic. The above idle vehicle relocation is integrated in the operating policy design of ridesharing system with transit transfers in the next section. Note that the decision variable $Y_{jm}$ is a relaxation from a binary variable to a continuous variable between 0 and 1, summed to a bounded integer by Eq. (8) to reduce the number of binary variables and reduce the complexity of the problem (Sayarshad and Chow, 2017). As discussed there, the queue intensity constraints (6) act as a piece-wise linearization which can be done using continuous variables as long as there is an additional constraint summing them to an integer. This computational reduction indeed differs from earlier queueing-based location problems like Marianov and Serra (2002) which assumed $Y_{jm}$ had to be binary and is considered a major contribution of Sayarshad and Chow (2017).

## 3. Proposed non-myopic dynamic vehicle dispatching and routing policy for ridesharing with transit transfers

The problem is modeled on a complete graph $G(N, E)$, where $N$ is a set of nodes and $E$ is a set of links. Each node represents either a transit station $i \in N_T$, a pick-up/drop-off point of ride requests $i \in N_P$, or a zone centroid where a vehicle may position itself $i \in N_Z$, $N = N_T \cup N_P \cup N_Z$. Travel time $t_{ij}$ is the shortest path travel time from node $i$ to node $j$. A ride request is characterized by its pick-up location, drop-off location, and desired pick-up time. For each node $i \in N$, the policy assumes request arrivals follow a Poisson process with arrival rate $\lambda_i$ with the set of all passengers denoted as $P$. Let $\mu_i, i \in N_Z$, denote the realized average service rate of vehicles of zone $i$, which is calculated over a time interval as total customers served divided by total in-vehicle time of the served customers.

For example, during a 10-minute relocation interval, there are three passenger drop-offs with pickup at zone $i$. If the time of vehicle drop-off minus time of vehicle pickup for those 3 passengers end up being 15 minutes, 10 minutes, and 20 minutes, the average $\mu_i$ for that 10-minute interval is 3 passengers/45 minutes. To avoid oscillating estimation of $\mu$ between relocation time intervals, we propose a three-step moving average method to adaptively learn $\mu$ over time. The service rate $\mu$ depends on the operator's dispatch and routing policy, arrival rate of customers and vehicles' positions.

The operation of the system is set as follows. The operator uses a fleet of homogeneous capacitated vehicles $V = \{v_1, v_2, \ldots, v_{|V|}\}$ to serve ride requests; the set of passengers assigned to a vehicle is $P_v$. A dispatching center makes decisions according to its operating policy for vehicle dispatching and route planning. Following past studies (Hyytiä et al., 2012; Sayarshad and Chow, 2015), we assume there is no time window constraints associated with the requests since it's a real time operation. All customers' requests are served so that operator costs can be quantified (subsequent studies can then use this information to design appropriate thresholds for rejecting requests, which varies by study region). Similar justification is made for user wait time; users do not cancel their requests in this study so that their wait time costs can be subsequently used to define appropriate thresholds during case by case implementations.

The operator determines dispatch and routing decisions for real-time trip requests (1) using operating vehicles only (direct trip) or by (2) using both operating vehicles (as last mile feeders) and fixed schedule Public Transport (PT) services. For the latter case, we assume there are *at most* two intermodal transfers for a customer's origin-destination trip. No transfer is allowed between two different rideshare operating vehicles as is the case in Liaw et al. (1996). Transfers impose a wait time on passengers. A customer's initial request is divided into three segments: a pre-transit trip (from origins to an entry station of PT system), an in-transit trip (from an entry stations to an exit stations), and a post-transit trip (from an exit station to a customer's destination). Each pre-transit trip or post-transit trip can be supported by either one individual vehicle or by foot. Travel time estimation of in-transit trips is based on the PT service schedules. For simplicity, the capacity constraint of PT vehicles is not considered in this study since rideshare users are assumed to make up only a small fraction of the general PT passengers.

The integrated problem and solution framework are highlighted in Fig. 2. The strategy is initiated by one of three different events: i) service option calculation for new arrival customer and vehicle dispatch; ii) idle vehicle relocation; or iii) vehicle dispatch when customers arrive at their exit transit stations. Each time a new passenger makes a request, the system runs a dynamic dispatch that considers the option of loading customers onto the transit system, using a proposed algorithm defined as P1 (this differs from the dispatch in Hyytiä et al. (2012), who only consider a single drop-off per passenger as opposed to selecting a station from candidates). Procedure P1 considers three possible options: a) 'rideshare only' (direct trip), b) 'rideshare-transit-walk (and vice versa)', and c) 'rideshare-transit-rideshare'. The 'rideshare-transit-rideshare' does not commit another vehicle for the post-transit trip immediately; if that option is chosen, the expected cost of using rideshare in the post-transit trip is incorporated into the options in P1. A passenger assigned this option would increase the demand at the exit station for the expected arrival time. When the passenger arrives at the exit station, the system runs another dispatch algorithm then as a 'rideshare only' to drop off the passenger to the final destination. The 'rideshare only' option is solved using an algorithm from Hyytiä et al. (2012). Note that marginal operational cost could also be incorporated for realistic applications to consider the tradeoff between user inconvenience and operational cost.

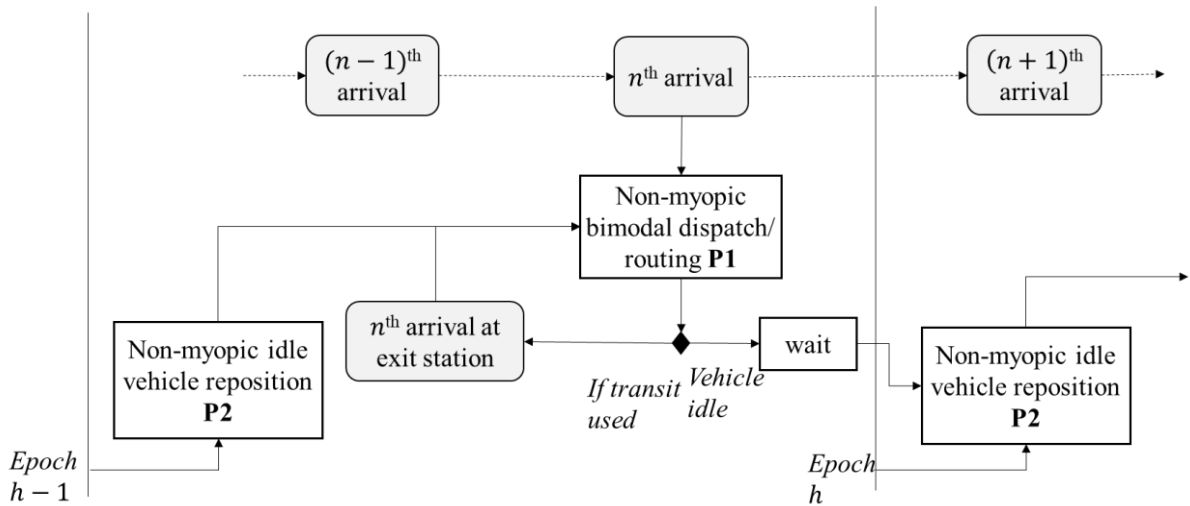

Fig. 2. Integrated strategy with functional components (rectangles) and initiating events (gray rounded rectangles)

We use the re-optimization-based insertion algorithm (Mosheiov, 1994) to solve the TSPPD to obtain $T(v,x)$ and $x$ for Eq. (3). This algorithm first finds a minimum-cost Hamiltonian tour for all drop-off locations of customers and then inserts pick-up locations one-by-one with cheapest cost in the Hamiltonian tour by satisfying precedence constraints and vehicle capacity. The Christofides heuristic (Christofides, 1976) is used to find an initial tour of the pickup locations. A 2-opt local search (Croes, 1958) is applied to improve the solution quality. As mentioned by Mosheiov (1994), the complexity of this algorithm depends on the method used to find the delivery tour which is $O(n^3)$. Other state-of-the-art TSP algorithms (Rego et al., 2011) can replace Christofides' heuristic. If we consider time-window constraints associated with customer's pickup and drop-off locations, Häme's algorithm (Häme, 2011) can be applied for large-scale problems. The use of k-nearest PT stations in Step 2 of P1 is a computational simplification, analogous to the use of k-shortest paths for online routing, to ensure that the vehicle dispatching runs in a timely manner.

**P1: Non-myopic vehicle dispatching and routing algorithm for rideshare with transit transfers**

1. Upon arrival of a new request $n$ traveling from $o$ to $d$, update positions and service statuses of every vehicle from the time of previous request $n-1$.
2. Compute a fastest option for request $n$ and its assigned vehicle among a number of nearby ones. Three options are considered for each nearby vehicle: 'rideshare only' (direct trip), 'rideshare-transit-walk (and vice versa)' and 'rideshare-transit-rideshare'. For rideshare only, travel time between origin and destination of a request is considered. For the other options, determine k-nearest PT stations to the origin and to the destination of a request. Travel time on a multimodal path sums up mode-specific (foot or vehicle) travel time from origin to entry station, waiting time at entry station due to transfer, in-transit travel time, waiting time at exit station, and travel time from exit station to destination. Travel time for the second ridesharing trip is approximated by considering the current system state. Choose the shortest multimodal travel time using one of k-nearest entry stations and one of the k-nearest exit stations.
    a. Determine the costs of the three service options: $C_r^n, C_{rtr}^n, C_{rtw}^n$ where $r$ is for rideshare only, $rtr$ is the rideshare-transit-rideshare option, and $rtw(wtr)$ is rideshare-transit-walk (or vice versa). The transit cost is the sum of waiting time (half of headway) plus travel time in transit system.
        i. For any option involving transit stations, consider each pair of k-nearest entry and exit stations, $(s1, s2)$.
        ii. For $r$: solve Eq. (2) for $(o, d)$ such that $[c(v, x) - c(v, x_t^v)]$ is minimized.
        iii. For $rtr$: solve Eq. (2) for each $(o, s1)$ and $(s2, d)$ such that the sum of costs $[c(v, x) - c(v, x_t^v)]$ for each trip plus transit cost is minimized.
        iv. For $rtw(wtr)$: solve Eq. (2) for each $(o, s1)$ (or $(s2, d)$) such that the sum of costs $[c(v, x) - c(v, x_t^v)]$ for the $(o, s1)$ or $(s2, d)$ trip plus transit and walking egress cost is minimized.
3. Update the pick-up or drop-off point of new request $n$ if that request uses a 'rideshare-transit-walk (and vice versa)' or 'rideshare-transit-rideshare option, respectively.
4. Update the new tour for that assigned vehicle.

Vehicles that have completed their service become idle. A relocation procedure, P2, is solved for all idle vehicles at the start of each relocation time interval (e.g. 10 minutes) to determine optimal zones to assign them. The non-myopic idle vehicle repositioning algorithm is based on Eqs. (3) – (16).

**P2: Non-myopic idle vehicle relocation policy for rideshare with transit transfers**
1. Upon start of a new idle vehicle relocation epoch $h$, identify the set of idle vehicles in fleet and their current zone locations and predicted values of $\lambda$ (including anticipated arrivals at exit stations that need rideshare service) for epoch $h$
2. Solve Eq. (3) to (16) using an MIP solver
3. For the vehicles assigned to a different zone, add the new location to the vehicle's tour.

The coordinates of zone centroids are calculated based on the center of gravity method (Thomas, 2007) and a 3-step moving average (Montgomery et al. 2008) proposed for this study, i.e. average over three preceding idle vehicle relocation epochs. The gravity center is calculated as shown in Eq. (17).

$$x_i^* = \frac{\sum_{s=1}^{m} \lambda_s x_s}{\sum_{s=1}^{m} \lambda_s}, \qquad y_i^* = \frac{\sum_{s=1}^{m} \lambda_s y_s}{\sum_{s=1}^{m} \lambda_s}, \forall s \in i, \text{for } i = 1, \dots, \overline{N} \qquad (17)$$

where $x_s$ and $y_s$ are the x-coordinate and y-coordinate of pick-up points $s$ within zone $i$. $\lambda_s$ is the arrival rate at $s$. Note that pick-up/drop-off points of vehicles correspond to the locations of customers' origins and destinations while zone centroids are separately designed for idle vehicle relocation. We propose updating the relocation zone centroids dynamically each epoch by considering the spatiotemporal variation of customer demand intensity.

The performance and characteristics of the underlying algorithms within P1 and P2 are discussed in Hyytiä et al. (2012) and Sayarshad and Chow (2017), respectively. P1's computational complexity is related to the underlying TSPPD problem to sequence a vehicle over a set of assigned passengers. The algorithm P1 is a heuristic that obtains a solution in polynomial time and has been shown to perform well. Algorithm P2 incorporates a MIP solver for a p-median problem. p-median problems are NP-complete; with fixed values of p they can be solved in polynomial time (Garey and Johnson, 1979; Owen and Daskin, 1998). The MIP solver used in this study is the mixed-integer linear programming solver 'intlinprog' from Matlab. For larger examples, alternative p-median heuristics can be employed, like the algorithm from Teitz and Bart (1968).

To summarize the contribution of this work, we propose a new system operating policy design that integrates a routing subproblem and an idle vehicle relocation subproblem in the presence of coexisting public transit service. It is not simply running two models together. Design issues include determining how to run them under different time frames (as highlighted in Fig. 2); design of separate online zone centroids for the relocation and stationing of idle vehicles; and integrating all that with public transit schedules over time. The presence of the public transit system, in particular, impacts the decisions of the subproblems; for example, idle vehicles end up endogenously locating closer to transit stations because of the influence of routing demands to transit stations, which we illustrate in the case study. It allows us to evaluate the impact of changes in subproblem objective parameters like weights between user and system performance in the routing on the overall performance of the integrated system. Compared to the earlier studies on developing dispatch and relocation models (Hyytiä et al., 2012; Sayarshad and Chow, 2017), the proposed method presents a number of new contributions summarized in Table 3.

Table 3 New Contributions Developed for Proposed Operational Strategy

| |
|---|
| We consider a dynamic multimodal door-to-door demand responsive transport service problem and design an online algorithm by considering both non-myopic vehicle dispatching and idle vehicle relocation integrated as shown in Fig. 2. |
| In a multimodal setting, the passenger path travel time involves estimating travel times for multiple legs (origin to station, station to station, station to destination) under different combinations. The use of k-nearest stations is needed to keep computational costs down since this would need to be operated in an online setting. |
| The dispatch operates off arrivals while the relocation operates off predefined epochs. Dispatch also occurs between stations and pickup/drop-off locations whereas relocation is conducted at a zonal level that needs to be determined/updated. To make the idle vehicle relocation algorithm work effectively in response to customer arrival pattern changes, we propose a dynamic scheme for zonal centroids updating over time. |
| The discrepancy in time and spatial units needs to be reconciled when combining the two algorithms, which means that we need to build up a whole simulation from scratch to run this strategy and test different parameters. The simulation incorporates the transit schedules to track when the passengers exit the stations. |
| The learning of the arrival and effective service rates are conducted in an online setting. As an integrated service, we can control other constraints like imposing maximum number of transfers (two) so in our case our solution allows for a range of four types of options provided to users: rideshare-only, RTR, RTW, and WTR. |

## 4. Numerical experiments

*4.1 Experimental design*

To test the effectiveness of the proposed operational strategy with all the system design changes, we conduct a series of numerical tests on a small instance. Two experiments are designed for this instance.

A) The first experiment considers rideshare only (no transit collaboration) to validate the methodology, compare its performance against varying degrees of myopic strategies, and to evaluate the sensitivity of the strategy to different parameters. The results of this experiment are shown in the Appendix.

B) The second experiment evaluates the sensitivity of the proposed strategy under different PT headways.

For benchmarking, we consider two alternative idle vehicle relocation policies aside from P2: (1) a "waiting policy" where an idle vehicle stays at their current position until a new dispatch is assigned to it; (2) a "busiest zone policy" where an idle vehicle moves to the busiest zone center (i.e. with highest customer arrival rate in average) with a probability of receiving at least one customer at the busiest zone higher than a threshold (Larsen et al., 2004). The probability of receiving $k$ new customers in a Poisson process with intensity $\lambda_j$ is calculated as Eq. (18).

$$P(X = k) = \frac{1}{k!}(\lambda_j t_{n_v n_s})^k e^{-\lambda_j t_{n_v n_s}}, \forall k = 0,1,... \qquad (18)$$

where $t_{n_v n_s}$ is travel time between node (zone centroid) $n_v$ and $n_s$. A user-defined threshold $\vartheta$ related to alternative 2 is specified to decide the relocation decision of idle vehicles as shown in Eq. (19).

$$P(X \geq 1) \geq \vartheta \qquad (19)$$

where $P(X \geq 1) = 1 - e^{-\lambda_j t_{ij}}$ is the probability of receiving at least one customer at the zone j. The threshold $\vartheta$ is a random variable drawn from the range of (0.5,1] to generate stochastic relocation decisions. The stochastic relocation decision reflects drivers' heterogeneous behavior in terms of willingness to reposition.

*4.2 Proposed simulation for evaluation*

We create a new discrete event simulation from scratch to test the proposed integrated operational strategy. Scheduled PT timetables and transit vehicle runs are implemented in the simulation to evaluate customers' experienced wait times at transit stations. A 2-hour customer arrival period following the Poisson distribution and uniformly distributed in the study area is considered. As passenger arrivals are based on stationary distributions, we do not need to run multiple simulation runs as long as we run the one simulation sufficiently long. The simulation is executed in MATLAB using a Dell Latitude E5470 laptop with win64 OS, Intel i5-6300U CPU, 2 Cores and 8GB memory. The test instance is publicly available on the following data library: https://github.com/BUILTNYU. The pseudocode of the simulation is described as follows.

1. Initialization: locate all vehicles at its initial depots. Mark each vehicle as 'available'. Initialize service rate $\mu_i = \mu_0$ for each zone $i = 1, ..., \overline{N}$. Set current simulation time $t = 0$, idle vehicle relocation warming-up period $T^{warm}$ and idle vehicle relocation epoch length $\Delta$. Initiate epoch $h = 0$.
2. Upon arrival of a new request $n$, compute a fastest option for the request $n$, and
   a. Update the system state up to the new request's arrival time:
      – Update the location of idle vehicles in transition (rebalancing during previous epochs) from the time of previous request. When an idle vehicle arrives at the relocated location, change the vehicle status as 'available', and remove the vehicle from in-transition idle vehicle list.
      – Update the state of the other vehicles from the time of previous request. If a vehicle drops off a customer at his/her entry transit station, compute the customers' waiting time for next train. For the dropped off customer, if he/she uses the 'rideshare-transit-rideshare' option, generate a new riding request for this customer with his/her exit transit station as the pickup point and the destination as the drop-off point. The desired pickup time is set as the expected arrival time at the corresponding exit station. Add the new request in the request list and sort the list by their pickup times; If the customer uses a 'rideshare-transit-walk' option, he/she takes the next train and gets to the destination by foot.
   b. Update *t* to the customer's arrival time. Run Algorithm P1 to dispatch a vehicle to pick up the new request. Note that allowing idle vehicles in repositioning to pick-up new customers is considered as an option for the system.

c. Idle vehicle relocation: If $t$ is greater than $\max(h\Delta, T^{\text{warm}})$, execute idle vehicle relocation. First, update the list of idle vehicles not in transition. Compute arrival rate $\lambda_i^h$, service rate $\mu_i^h$ and zone center $(x_i^{*h}, y_i^{*h})$ for each zone $i$ based on the 3-step moving average method. Compute $W_{ij}$ based on the idle vehicle relocation policy. Update $h := h + 1$.
d. Repeat until all new requests are served.
3. Advance the clock through the remaining tours until all customers are served.

## 4.3 Test instance

We consider a region on a plane bounded by (-10,-10)×(10,10), representing a 20 km × 20 km area shown in Fig. 3. The entire region is divided into 16 identical relocation zones with each zone of 5-km in length and width. Customers are assumed to arrive randomly following the Poisson process. The rideshare operator uses a fleet of identical capacitated vehicles for real-time MoD service requests. All vehicles are initiated at the center depot (0, 0) (a warmup period is used to position the vehicles more naturally).

A simple transit network is overlaid with the solid blue lines that includes 89 transit stations. Transit routes are set to operate along both directions with pre-defined headways between transit vehicles.

The reference parameters used in the simulation are listed in Table 4. Two performance metrics are used to measure the performance of the proposed methodology: mean travel time of vehicles (system operating cost) and mean journey time per passenger (customers' inconvenience). The journey time of customers is the difference between their arrival time and drop-off time. The number of simulation runs does not influence the simulation output when same initial condition is applied. Other measures can also be considered—e.g. average vehicle occupancy, passenger-km/veh, passenger-hour/veh, and rerouting distance increment—but for conciseness we leave that work for subsequent implementation studies.

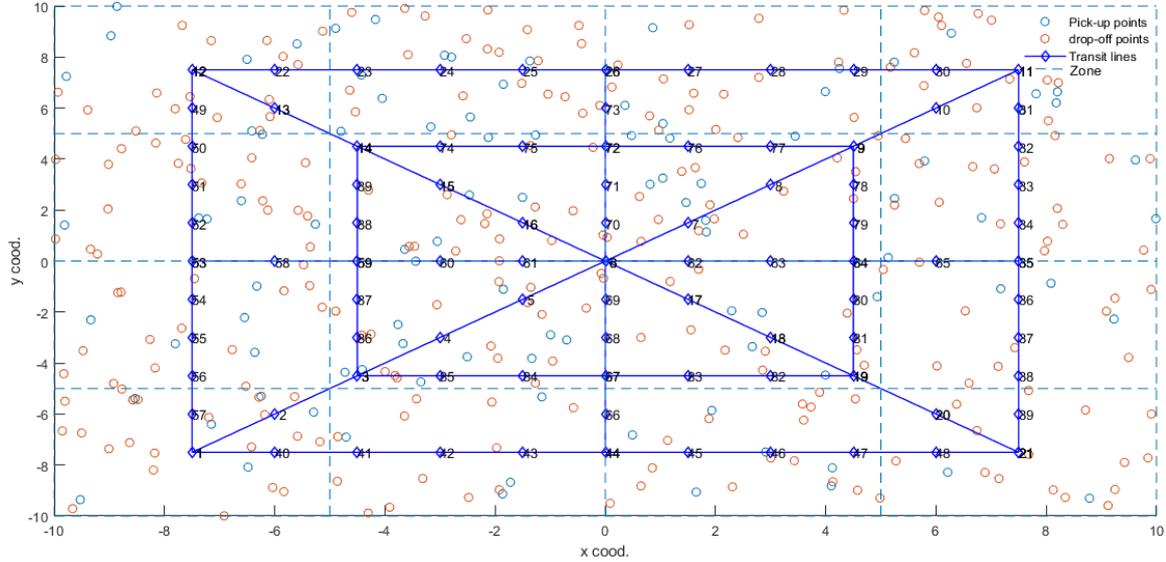

Fig. 3. Test instance for the bimodal dynamic dial-a-ride problem

Table 4 Reference parameter settings for the simulation evaluation

| | |
|---|---|
| Fleet size (vehicle) | 40 |
| Number of zones ($\bar{N}$) | 16 |
| Capacity of vehicle $q$ (customers/vehicle) | 4 |
| Vehicle speed (km/hr) | 36 |
| Train speed (km/hr) | 80 |
| Walk speed (km/hr) | 5 |

| | Idle vehicle relocation interval (minute) | 10 |
|---|---|---|
| | Warming up period (minute) | 10 |
| | k-nearest entry (exit) stations considered* | 4 |
| | $\gamma$ (Eq. (3)) | 0.5 |
| | $\rho_{\eta j m}$ | $\eta = 0.95$, b=0 for all $j, m$ |
| | Customer arrival period | 2 hours |

Remark: A total of $k^2 = 16$ paths are searched when determining rideshare-transit option for each vehicle

We study the influence of the key parameters and customers' arrival intensity on the performance of the system and validate the proposed methodology. We vary the values of these parameters and define the test scenarios as follows.

- Customer arrival intensity: two situations corresponding to low and high arrival intensity are tested: $\lambda = 100$ and 400 customers/hour.
- $\beta$ is related to the degree of look-ahead in vehicle dispatching, which needs to be calibrated in order to find an adequate value (Hyytiä et al., 2012). We test three sets of data points of $\beta$ to limit computational cost to find a good value of $\beta$, i.e. $\frac{k}{\bar{T}(v,x)}, \frac{0.1k}{\bar{T}(v,x)}, \frac{0.01k}{\bar{T}(v,x)}, k = 1, ..., 10$.
- $\theta$ is a scaling parameter to convert idle vehicle relocation cost in the objective function of the mixed-integer linear programing problem (P2). We set idle vehicle relocation cost $r_{ij}$ related to rebalancing travel time from zone $i$ to $j$, i.e, $r_{ij} = t_{ij}$.
- Idle vehicle relocation policy: We test four idle vehicle relocation policies: a) waiting policy (no rebalancing); b) busiest zone policy; c) myopic idle vehicle relocation policy by relaxing the queueing delay constraint (Eq. (6)); d) non-myopic idle vehicle relocation policy of P2.
- Rideshare only versus rideshare-transit options: we vary transit vehicle headways of 5, 10 and 20 minutes to evaluate the benefit of rideshare-transit options. 'Rideshare only' (no transit-rideshare collaboration) is the benchmark to evaluate the benefit of transit-rideshare collaboration.
- $\rho_{\eta j m}$ is a user-defined parameter. The idle vehicle relocation interval is set up to ensure idle vehicles have sufficient time to reach their customers (Sayarshad and Chow, 2017). The warming up period setting is necessary to avoid unnecessary rebalancing operations at the beginning of the simulation.

*4.4 Results*

The results of the influence of different system design parameters on the system with rideshare only are reported in the Appendix.

Having established the baseline performance of the nonmyopic algorithms in the test instance, we assess the benefit of introducing transit transfers on the system performance with different headways as shown in Table 5.

Table 5 Benefits of the system with rideshare-transit options

| $\lambda$ | Headway of transit vehicles (minutes) | Passenger waiting time (minute) | | Mean passenger journey time (minute) | Mean vehicle travel time (minute) | Rideshare-transit option | | | |
|---|---|---|---|---|---|---|---|---|---|
| | | Mean | Max | | | R | WTR | RTW | RTR |
| 100 | - | 11.6 | 39.5 | 34.5 | 90.6 | 100% | - | - | - |
| | 5 | 6.9 | 29.3 | 35.1(+1.9%) | 48.0(-47.0%) | 19.5% | 41.5% | 36.0% | 3.0% |
| | 10 | 7.0 | 26.6 | 36.4(+5.5%) | 49.9(-44.9%) | 21.0% | 38.5% | 35.5% | 5.0% |
| | 20 | 7.0 | 32.3 | 38.6(+11.9%) | 52.9(-41.6%) | 31.5% | 31.0% | 32.0% | 5.5% |
| 400 | - | 89.8 | 317.0 | 126.7 | 378.4 | 100% | - | - | - |

| | | | | | | | | |
|---|---|---|---|---|---|---|---|---|
| 5 | 21.0 | 94.1 | 50.0(-60.5%) | 171.9(-54.6%) | 39.4% | 27.4% | 27.1% | 6.1% |
| 10 | 23.1 | 102.0 | 52.6(-58.5%) | 177.9(-53.0%) | 44.9% | 26.0% | 24.4% | 4.8% |
| 20 | 28.6 | 118.6 | 61.1(-51.8%) | 194.3(-48.6%) | 52.4% | 23.6% | 21.4% | 2.6% |

Remark: 1. R: rideshare only, RTW/WTR: rideshare+transit+walk / walk+transit+rideshare, RTR: rideshare+transit+rideshare. 2. Passenger waiting time is a passenger's total waiting time at pick-up points for rideshare vehicles. 3. Idle vehicles in transition to its assigned zone are allowed to pick up new customers.

The smaller the transit headway, the higher the ratio of customers with transit transfer options. The rideshare-transit cooperation can effectively reduce operation cost with significantly lower vehicle traveled miles (-47.0% for $\lambda = 100$ and -54.6% for $\lambda = 400$ in the case of 5-minute headway of transit) and lower user journey time (-60.5% for $\lambda = 400$ scenario with a 5-minute headway of transit). This is because for high arrival customer intensity there is a synergistic effect of dropping off/picking up customers at the same stations. It is like the concept of meeting points to enhance the efficiency of first/last mile pick up or delivery in a ridesharing system (Stiglic et al., 2015). For the rideshare-transit option, WTR/RTW are the main adopted options. It represents 63%-77.5% for the scenario of $\lambda = 100$ and 45%-54.5% for the scenario of $\lambda = 400$. On the other hand, the ratio of RTR is marginal (around 6% or less) due to its higher operating cost for the first and last mile connecting rides.

By adding PT to create new service options, the improvement vastly outperforms the improvement seen from only adopting non-myopic algorithms, with ***up to 60% reduction in user journey times for "non-myopic + PT" on top of the 2-3% reduction from just "non-myopic"***. These tests confirm that integrating rideshare with PT holds tremendous potential. A case study is needed to evaluate the performance of the same algorithm using realistic demand, transit, and network data.

## 5. New York City and Long Island Railroad case study

The case study is designed to answer the following research questions using realistic travel demand data:
- How much better can a system with transit transfers outperform rideshare-only system when operating non-myopic algorithms under different traffic conditions?
- Under what conditions is rideshare with integrated transit preferred, and within those conditions when are RTW/WTR preferred over RTR?
- How much does the effective service capacity increase under the proposed strategy?
- How do we use the algorithm to plan for service expansions?

*5.1 Data*

The NYC metropolitan region focusing on Long Island is shown in Fig. 4. This setting is ideal for the proposed policy because the distance is too far for door-to-door rideshare service. Driving trips from Long Island to NYC typically can take 1 to 3 hours depending on origin and time of day. For the demand data, we use 2010-2011 Regional Household Travel Survey of New York metropolitan area conducted by the NYMTC Metropolitan Planning Organization (NYMTC, 2018). The data shown in Fig. 4 corresponds to all the trips made between 7:00-9:00AM regardless of their mode; the experiment assumes all these trips are potential shared taxi pickups and drop-offs to compare between "rideshare only" to the proposed strategy. For LIRR service, we use a frequency of 20-minute headway for all the stations.

We exclude Bronx and Staten island from the study area as they are not directly accessible via LIRR. The studied area is divided into 72 zones (i.e. $\bar{N}$ with $N_z$ is the zone centroid of zone z) based on the State Legislative Districts and five counties: Suffolk (LI), Nassau (LI), Queens (NYC), Kings (NYC-Brooklyn), and New York (NYC-Manhattan). The OD demand is aggregated at the "Transportation Analysis Zones (TAZ)" level. This zoning system shown in Fig. 5 has 2096 zones for NYC and Long Island. For context, the 2010 population of the three NYC counties (New York, Queens, Kings) is 6.3M and for the two Long Island counties (Suffolk, Nassau) it is 2.8M.

Customer arrival intensity during morning peak hours varies considerably, especially for Queens and Brooklyn. Trips are made between all the counties; assuming there is a rideshare service for these five counties, the fleet will have to split its time between serving direct trips for some (primarily NYC-to-NYC county trips) and providing last mile service for other multimodal trips. A summary of the customer arrival patterns into the system (i.e. departure times) from each of the five counties in the Household Travel Survey samples is shown in Fig. 6.

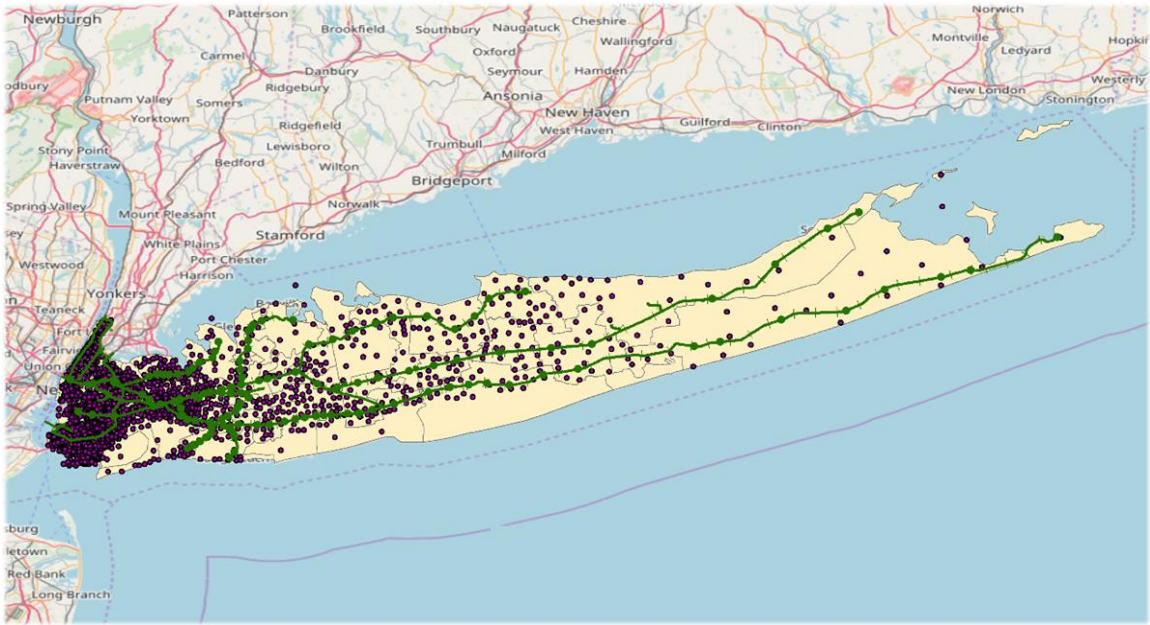

Fig. 4. Pickup and drop-off points from 2010/2011 Household Travel Survey in New York City and Long Island in 7:00 – 9:00 a.m. The green line is the Long Island Rail Road (NYS Railroad Lines shapefiles: https://gis.ny.gov/gisdata/inventories/member.cfm?OrganizationID=539)

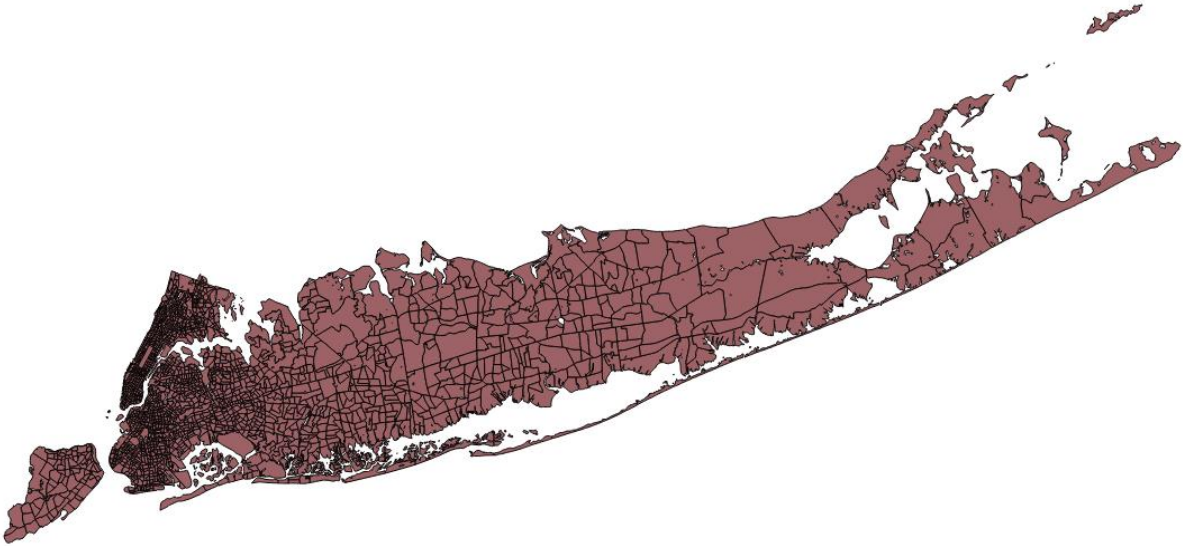

Fig. 5. "Transportation Analysis Zones (TAZ)" system for the NYC and Long Island (NYMTC, 2000. Traffic Analysis Zone shapefiles, 2010/2011 Regional Household Travel Survey, available by request)

Parameters for the simulation and models are summarized in Table 6. Vehicle speed considers congested travel times based on an average speed estimated by sampling trips using Google Maps in Jan. 2018 during morning peak hours in the study area. We test three scenarios with increasing fleet size, i.e. 720, 1440, and 2160 vehicles, which correspond to 10, 20 and 30 vehicles initiated in each zone. The key parameters $\beta$ and $\theta$ are calibrated to be effective for the proposed methodology.

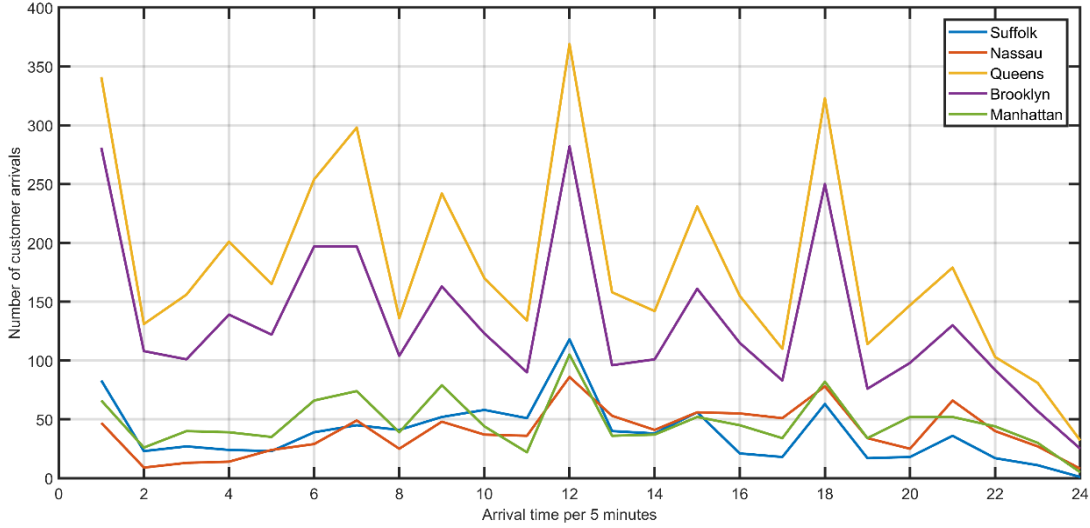

Fig. 6. Customer arrival times over counties in NYC and Long Island during 7:00 – 9:00 a.m.

Table 6 System characteristic and parameter settings for NYC and LIC case study

| | | | |
|---|---|---|---|
| Number of customers | 10572 | $\beta$ | $4/\bar{T}(v,x)$ |
| Number of zones | 72 | $\gamma$ | 0.5 |
| Fleet size | 720/1440/2160 | $\rho_\eta$ | $\eta = 0.95$, b=0 |
| Capacity of vehicles | 4 pers./veh. | Idle vehicle relocation interval | 15 min. |
| Walking speed | 5 km/hour | Warming up period | 30 min. |
| Vehicle speed | 29.4 km/hour | Headway of train | 20 min. |
| Number of transit stations | 124 | Simulation time | 2 hours |
| k-nearest entry (exit) stations considered | 4 | | |

Remarks: 1. $\bar{T}(v,x)$ is mean vehicle travel time without considering transit-rideshare cooperation.

*5.2 Parameter calibration*

We first calibrate the parameters. To make the proposed method effective, we decide a warm-up period to avoid unnecessary relocation at the beginning period of the service. As the fleet size is large in the application, we consider $n = 20$ nearest vehicles for new requests for the non-myopic vehicle dispatching policy (Eq. 1). All vehicles are initially located at each zone center instead of a centralized depot.

We set the idle vehicle relocation interval as 15 minutes as it is an approximate travel time to reach a neighbor zone. The warming-up period is tested up to 40 minutes to assess its impact on the performance of the system. As shown in Table 7, using a warm-up time of 40 minutes produces the most effective results in terms of customers' inconvenience and operation cost.

To ensure an effective relocation policy, we calibrate $\theta$ over a range of values between 0 and 80. The result is shown in Table 8. A $\theta = 20$ has the best performance over different fleet sizes. When comparing the performance with that of the benchmark, the proposed methodology reduces considerable operation cost with higher average journey time per passenger in highly congested cases, i.e. 10 vehicles per zone. When increasing the fleet size, the proposed idle vehicle relocation becomes less effective.

Table 7 The influence of warming-up time on the performance of the system with rideshare only

| Warming-up time (minute) | Number of vehicles per zone | | | | | | | | |
|---|---|---|---|---|---|---|---|---|---|
| | 10 | | | 20 | | | 30 | | |
| | WT | JT | VTL | WT | JT | VTL | WT | JT | VTL |
| 10 | 77.2 | 151.1 | 462.2 | 13.9 | 61.4 | 214.4 | 10.7 | 53.4 | 157.7 |
| 20 | 79.9 | 152.2 | 460.8 | 14.7 | 62.3 | 218.0 | 10.7 | 53.2 | 156.8 |
| 30 | 79.9 | 152.2 | 460.8 | 14.7 | 62.3 | 218.0 | 10.7 | 53.2 | 156.8 |
| **40** | **62.1** | **133.0** | **440.5** | **13.2** | **60.7** | **210.8** | **8.7** | **51.7** | **146.7** |

Remark: 1. WT: Mean passenger waiting time, JT: Mean passenger journey time, VTL: Mean vehicle travel time. 2. Passenger waiting time is a passenger's total waiting time at pick-up points for rideshare vehicles. 3. Measured in minutes.

Table 8 The calibration of $\theta$ for the idle vehicle relocation policy of the NYC case study

| $\theta$ | Number of vehicles per zone | | | | | | | | |
|---|---|---|---|---|---|---|---|---|---|
| | 10 | | | 20 | | | 30 | | |
| | WT | JT | VTL | WT | JT | VTL | WT | JT | VTL |
| Benchmark | 58.7 | 123.2 | 467.9 | 13.2 | 57.2 | 220.8 | 8.0 | 47.4 | 147.5 |
| 2 | 83.5 | 156.3 | 470.2 | 25.3 | 75.4 | 243.2 | 12.6 | 55.9 | 164.2 |
| 4 | 75.9 | 149.5 | 460.9 | 22.6 | 73.1 | 238.2 | 9.6 | 52.4 | 152.7 |
| **20** | **62.1** | **133.0** | **440.5** | **13.2** | **60.7** | **210.8** | **8.7** | **51.7** | **146.7** |
| 40 | 62.0 | 132.6 | 441.9 | 13.2 | 60.7 | 210.8 | 8.7 | 51.7 | 146.7 |
| 60 | 61.6 | 133.0 | 440.2 | 13.2 | 60.7 | 210.8 | 8.7 | 51.7 | 146.7 |
| 80 | 61.6 | 133.0 | 440.2 | 13.2 | 60.7 | 210.8 | 8.7 | 51.7 | 146.7 |

Remark: 1. Benchmark is the system without non-myopic vehicle dispatching and idle vehicle relocation. 2. WT: Mean passenger waiting time, JT: Mean passenger journey time, VTL: Mean vehicle travel time. 3. Passenger waiting time is a passenger's total waiting time at pick-up points for rideshare vehicles. 4. Measured in minutes.

*5.3 Results: Increase in effective service capacity*

We analyze the benefit of transit and rideshare collaboration with respect to different fleet sizes from 10 to 30 vehicles per zone. Table 9 reports the performance of the system with rideshare only and the proposed strategy, where reductions in costs are desired.

Table 9 Benefit of the system with transit-rideshare options compared to that of rideshare only

| Number of vehicles per zone | System with rideshare only | | | | System with rideshare-transit options | | | |
|---|---|---|---|---|---|---|---|---|
| | WT | JT | VTL | Comp. time (min.) | WT | JT | VTL | Comp. time (min.) |
| 10 | 62.1 | 133.0 | 440.5 | 77.0 | 19.4(-68.7%) | 60.8(-54.3%) | 175.2(-60.2%) | 698.0 |
| 20 | 13.2 | 60.7 | 210.8 | 86.6 | 5.6(-57.6%) | 41.2(-32.1%) | 76.5(-63.7%) | 636.9 |
| 30 | 8.7 | 51.7 | 146.7 | 90.1 | 5.5(-37.0%) | 41.6(-19.6%) | 55.7(-62.1%) | 678.4 |

Remark: 1. WT: Mean passenger waiting time, JT: Mean passenger journey time, VTL: Mean vehicle travel time. 2. Passenger waiting time is a passenger's total waiting time at pick-up points for rideshare vehicles. 3. The computational time is the total simulation time of each case study. The average computational time for solving a single P1 for a new request is 2.3 sec. The average computational time for solving a P2 is 34.7 sec. Both are based on the case of 20 vehicles/zone.

For the system with rideshare only, when increasing the fleet size from 10 to 20 vehicles per zone, the mean passenger journey time and mean vehicle travel time decrease 54.3% and 52.2%, respectively. When further increasing the fleet size to 30 vehicles per zone, its marginal benefit in reducing passenger journey time reduces to -14.8% only. The marginal passenger journey time reduction is not proportional to that of fleet size increase. When assessing the benefit from rideshare and transit collaboration, passengers' journey time and system operating cost are substantially reduced. For the scenarios of 10 vehicles per zone, the average passenger journey time is reduced by 54.3%. The mean vehicle travel time is reduced 60.2%. If we change the fleet size to 20 vehicles per zone, the mean journey time reduces to 41.2 minutes. ***Compared to the system with rideshare only, the benefit of rideshare-transit option is still substantial: -32.1% in mean passenger journey time and -63.7.1% in mean vehicle travel time for the scenario of 20 vehicles per zone.***

When further increasing the fleet size to 30 vehicles per zone, the benefit in reducing mean passenger journey time is still significant (-19.6%) and the mean vehicle travel time is cut by 62.1%. These are significant savings: increasing fleet size from 10 to 30 vehicles per zone reduces average rideshare-only user journey time (waiting and riding time) from 133 minutes to 51.7 minutes while reducing average vehicle trip length from 440.5 minutes to 146.7 minutes. Having the transit option further reduces those numbers: 133.0 to 60.8 minutes (for 10 vehicles/zone) and 51.7 to 41.6 minutes (for 30 vehicles/zone) for the user journey time, and 440.5 to 175.2, and 146.7 to 55.7 for the average vehicle trip length.

We conduct several other comparisons on spatial distribution under the proposed strategy for the fleet size of 1440 vehicles. Fig. 7 reports the evolution of $\lambda$, $\mu$ and average number of idle vehicles per zone over time for the system with rideshare only and the system with transit transfers. Average vehicle service rate and average number of idle vehicles per zone are much higher for the system of rideshare with transit transfers. By having the transit option, the fleet is spending much less time making long distance trips serving customers. This is indicated by Fig. 8 and 9, where the average number of idle vehicles per zone over each 15-minute interval is much lower when there's only rideshare compared to the case with transit access, for the same fleet size.

Intuition suggests having RTR trips might increase user journey time due to added transfers. This is not the case, however. By having the rideshare service focus on providing first/last mile trips instead of direct door-to-door trips, the fleet of vehicles are made available more often. This means that the presence of the PT network provides the MoD service with an effective boost in capacity. This capacity boost is equivalent to a multiplier of $\frac{12.48-2.47}{2.47} = 4.05$ due to decreasing the average trip length served by each vehicle (since they don't have to go door-to-door) by $\frac{34.159-47.466}{47.466} = 28\%$. ***This means adding shared transit to the rideshare service can increase effective available vehicle capacity by 4 times while reducing trip length by only 28%.***

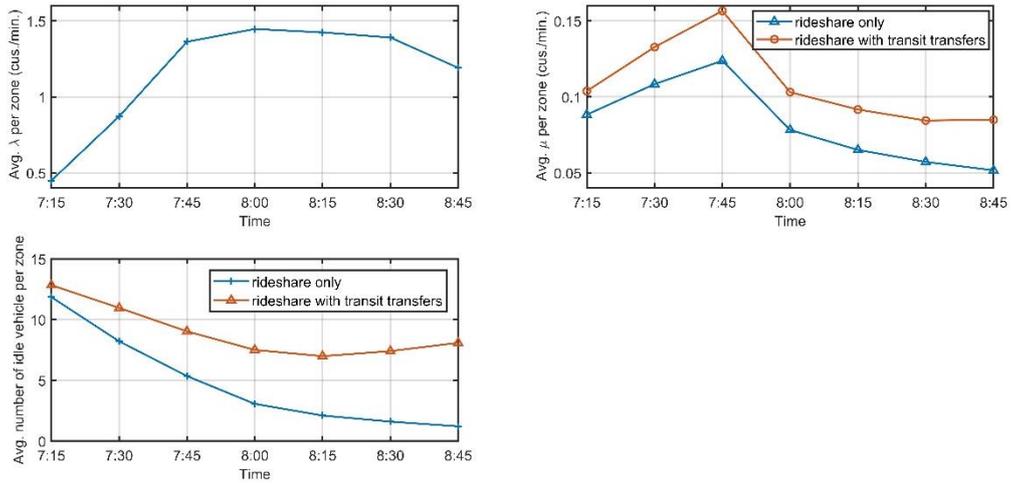

Fig. 7. Evolution of $\lambda$, $\mu$ and average number of idle vehicles per zone over time for NYC case studies (fleet size=1440 vehicles)

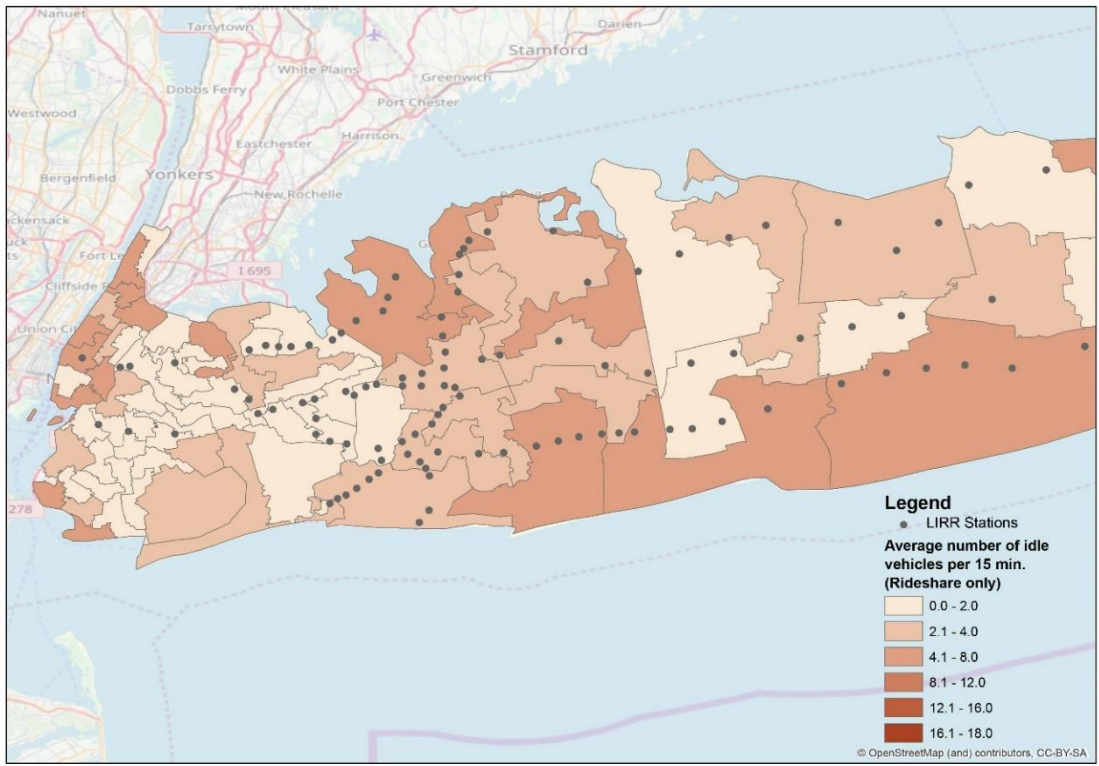

Fig. 8. Average number of idle vehicles per zone per 15 minutes (system with rideshare only)

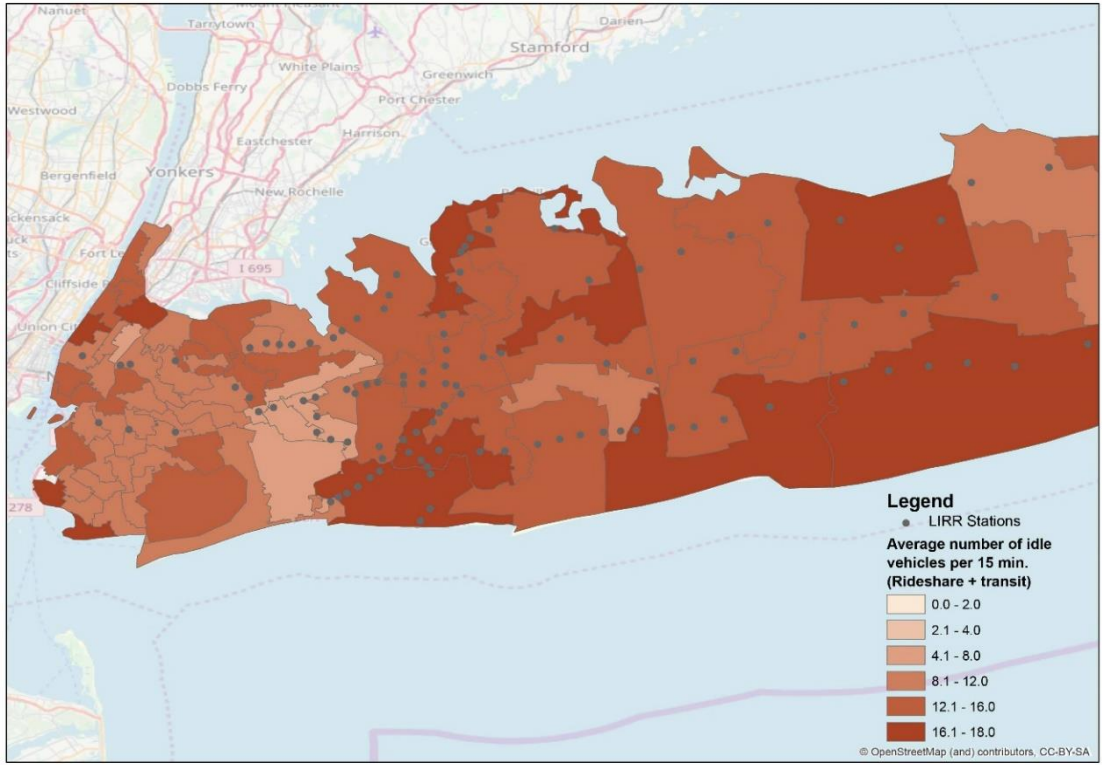

Fig. 9. Average number of idle vehicles per zone per 15 minutes (system with rideshare + transit)

In Fig. 10, the rideshare only scenario leads to very few available vehicles for rebalancing during the 2-hour simulation period since they end up making long trips. Instead, the rebalancing mostly occurs in Manhattan. Fig. 11 shows that with having transit capacity accounted for, there is more capacity to work with when considering rebalancing needs as the average trip served is also significantly shortened. The average passenger trip per vehicle is doubled from 0.85 to 1.61 (passengers/vehicle/hour) for the scenario of 10 vehicles per zone. The gains in terms of average passenger trip per vehicle become +39.7% (+28.8%) for the scenarios of 20 (30) vehicles per zone.

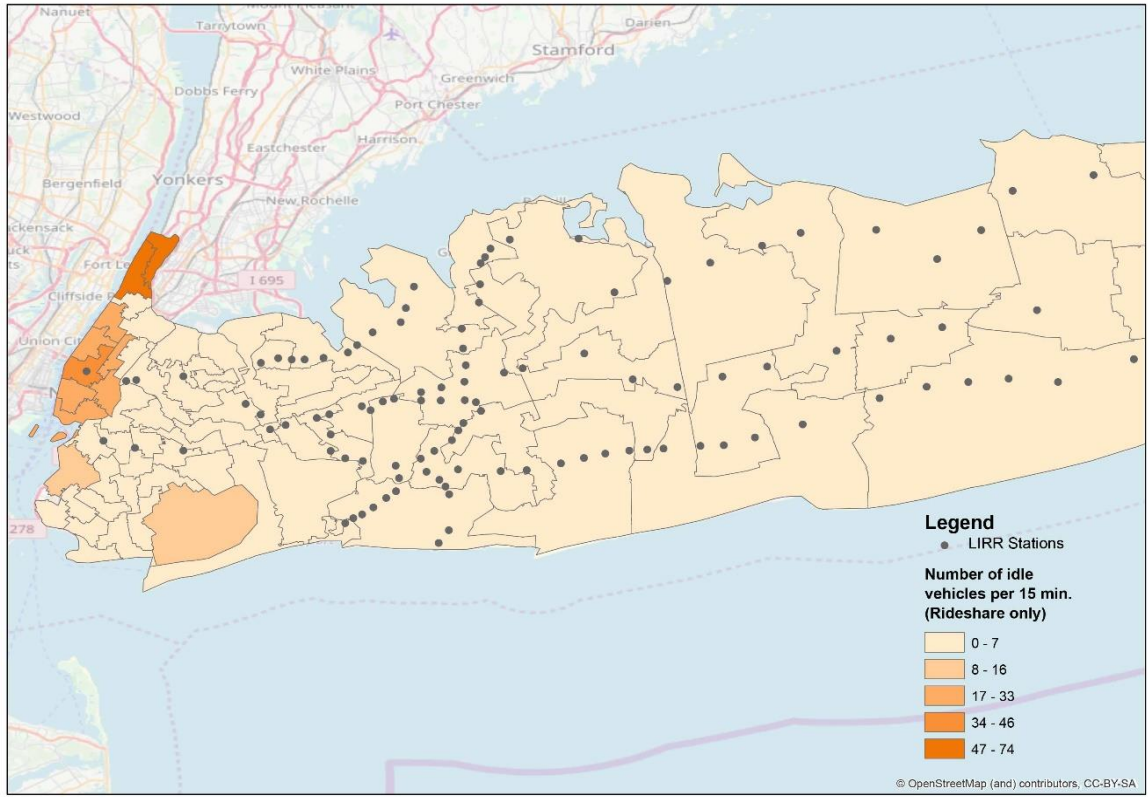

Fig. 10. Number of rebalanced vehicles in zones (system with rideshare only)

***The impact of these shorter trips is visualized in Figs. 12 and 13.*** In Fig. 12, the rideshare only trips tend to result in much longer lines overall; in Fig. 13, the availability of transit service capacity reduces the collective lengths of trips.

*5.4. Results: Demand distributions*

Table 10 reports the ratio of customers using rideshare only and both rideshare and transit service. There is around 57.9 – 62.9% using rideshare only option. The RTW option takes up 31.0 – 36.8% for the scenarios. The WTR and RTR represent less than 5% each. The share of people taking the ridesharing as a last mile (WTR) is small compared to RTW because the morning time period studied has most trips coming from Long Island to NYC and not the other way around. As a result, the first-mile in Long Island benefits much more from having rideshare access than the last mile in Manhattan. This split reflects the demand patterns and the underlying PT network structure. As shown in the synthetic example earlier the distribution can be very different.

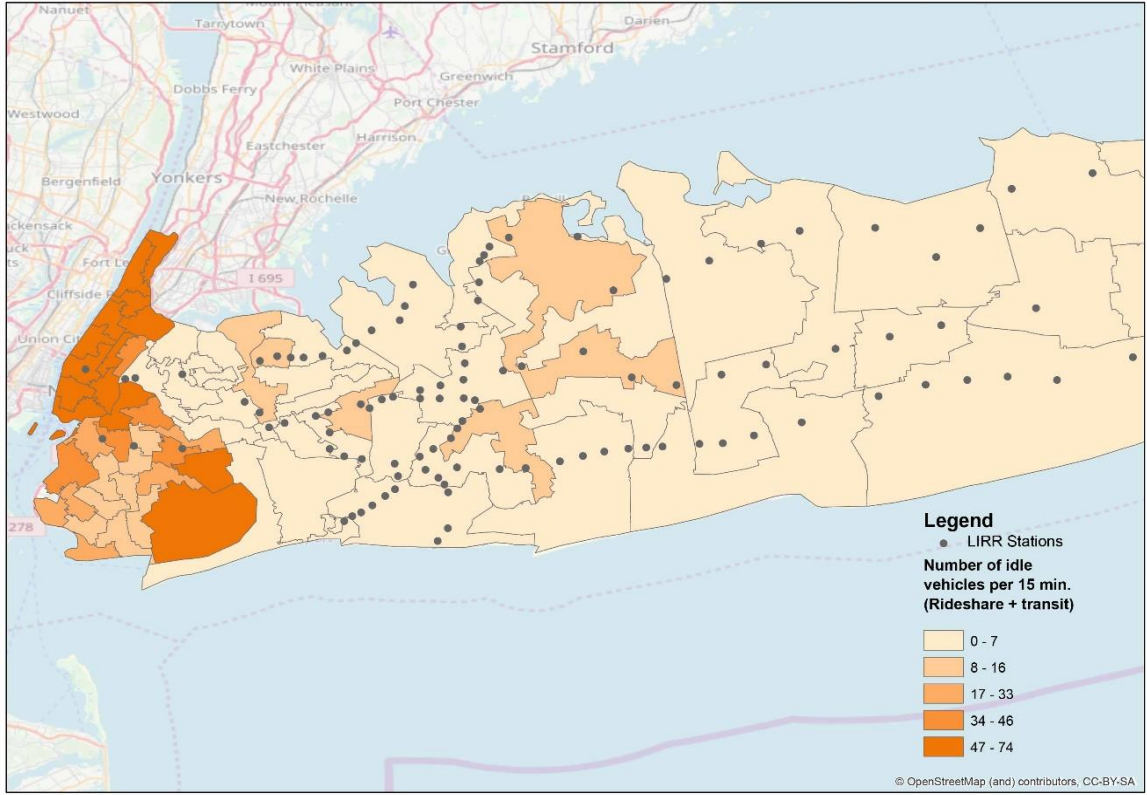

Fig. 11. Number of rebalanced vehicles in zones (system with rideshare + transit)

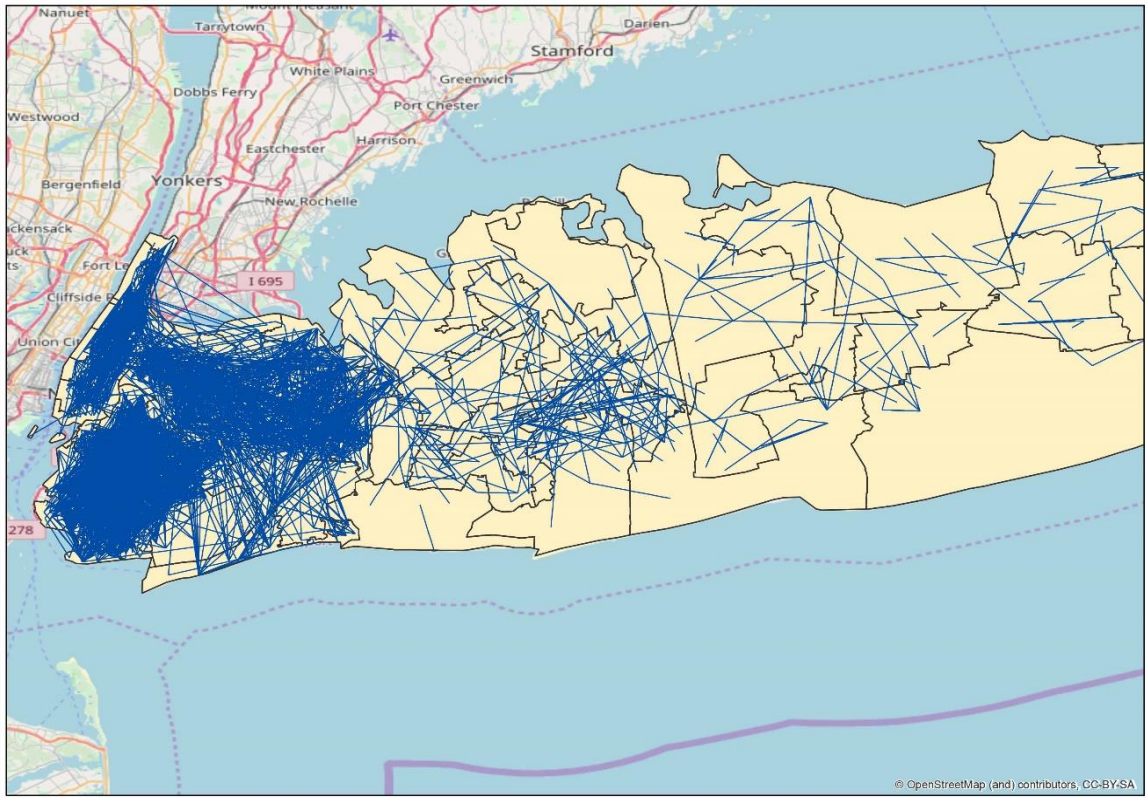

Fig. 12. Spatial distribution of trips using rideshare only

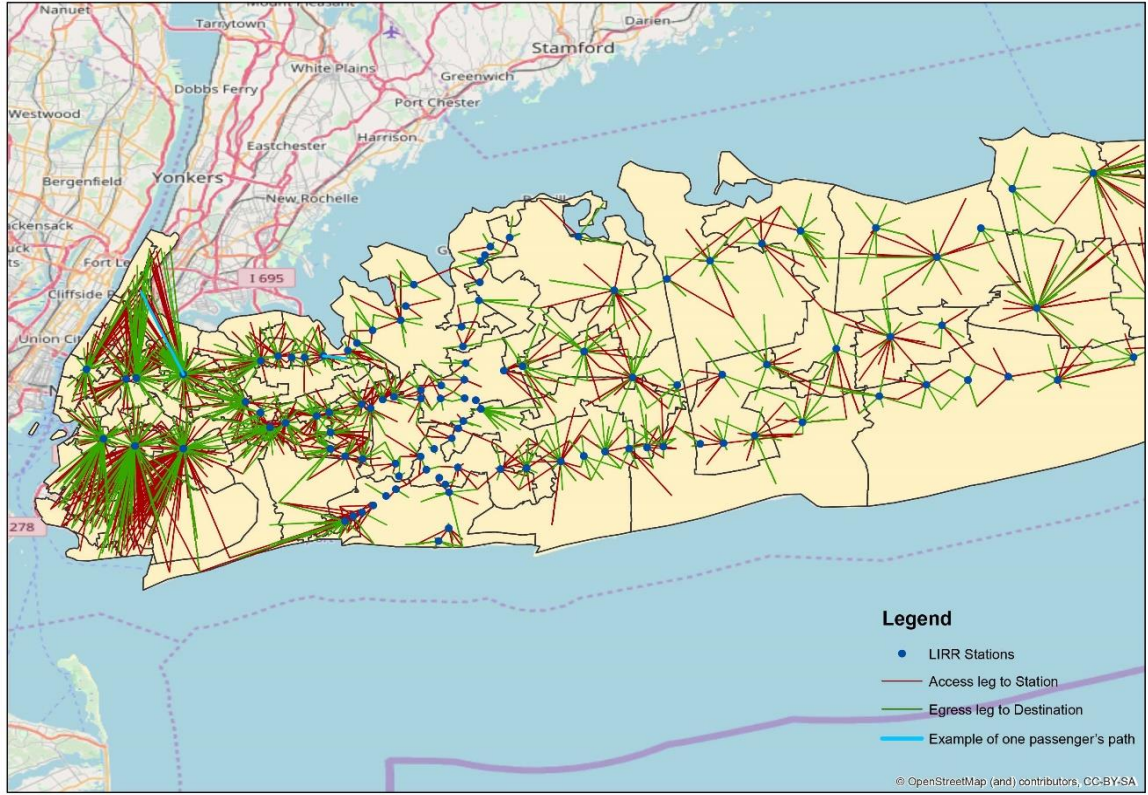

Fig. 13. Spatial distribution of trips using rideshare-transit options

Table 10 Ratio of different rideshare-transit options customers for the NYC-LI case study

| Number of vehicles per zone | WTR | RTW | RTR | R |
|---|---|---|---|---|
| 10 | 4.7% | 31.0% | 4.4% | 59.9% |
| 20 | 0.3% | 34.1% | 2.7% | 62.9% |
| 30 | 1.0% | 36.8% | 4.4% | 57.9% |

Remark: W: Walk, T: Transit, R: Rideshare

In Table 11a, the percentage of WTR, RTW and RTR for each of the OD counties are shown. The trips between Brooklyn and Manhattan do not use transit because the LIRR is not designed to provide coverage between those two counties. On the other hand, the transit-rideshare option makes sense for LI commuters going to NYC. These results show that there are settings where the proposed strategy is highly warranted.

Table 11a Percentage (%) of transit-rideshare service usage for different OD counties

| From/To | Suffolk | Nassau | Queens | Brooklyn | Manhattan |
|---|---|---|---|---|---|
| Suffolk | 73.8 | 95.8 | 100.0 | 100.0 | 100.0 |
| Nassau | 82.1 | 67.6 | 100.0 | 100.0 | 100.0 |
| Queens | 100.0 | 93.9 | 40.3 | 52.2 | 59.0 |
| Brooklyn | 100.0 | 100.0 | 100.0 | 3.1 | 62.5 |
| Manhattan | NA | 100.0 | 72.5 | 85.7 | 1.5 |

Table 11b Average journey time of systems with/without transit option for different OD counties

| From/To | Suffolk | | Nassau | | Queens | | Brooklyn | | Manhattan | |
|---|---|---|---|---|---|---|---|---|---|---|
| | R | R+T | R | R+T | R | R+T | R | R+T | R | R+T |
| Suffolk | 151 | 96.1 -36.4% | 182.6 | 94.7 -48.1% | 248.7 | 104.3 -58.1% | 258.2 | 159.3 -38.3% | 284.7 | 238.2 -16.3% |
| Nassau | 92.1 | 64.4 -30.1% | 68.5 | 52.9 -22.8% | 112.3 | 86.6 -22.9% | 130.9 | 124.2 -5.1% | 106.4 | 149.6 +40.6% |
| Queens | 189 | 89.5 -52.6% | 96.6 | 64.2 -33.5% | 58.3 | 36.6 -37.2% | 111.4 | 76.6 -31.2% | 79.1 | 67.7 -14.4% |
| Brooklyn | 218 | 124.7 -42.8% | 131.4 | 81.8 -37.7% | 65.5 | 69.6 +6.3% | 23.9 | 20.3 -15.1% | 55.5 | 74.8 +34.8% |
| Manhattan | NA | NA | 110.3 | 82.5 -25.2% | 56 | 59.7 +6.6% | 63.6 | 74.6 -15.1% | 15.4 | 14.9 -3.2% |

Remark: 1. R: system with ride share only. R+T: system with transit option. 2. Measured in minutes.

Cumulative probability distributions of trips with rideshare only and those with transit option under the 20 vehicles per zone setting are reported in Fig. 14. The proposed strategy reduces passengers' journey time for a certain distance range (in this example, between 30 to 300 minutes) for which having transit service becomes more advantageous compared to rideshare only.

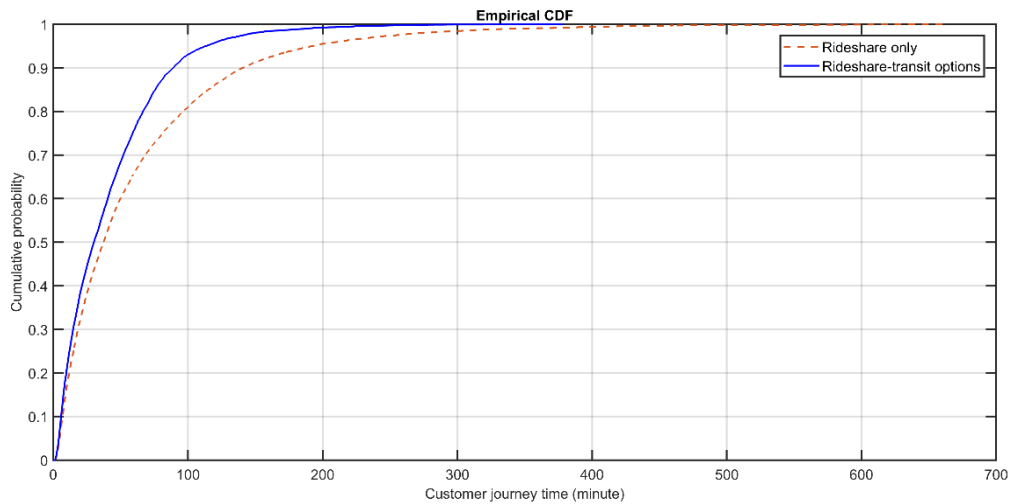

Fig. 14. Example of cumulative probability distributions of customer journey times for systems with rideshare only and with bimodal options (fleet size of 20 vehicles per zone)

*5.5. Results: Service coverage expansion decision support*

One more analysis is conducted for service coverage. Suppose a ridesharing service had to consider between expanding to either Suffolk County or to Nassau County. A simulation of demands coming from or going to each county as a separate scenario is conducted. The demand between NYC and the two counties is shown in Table 12. We test three scenarios with 500, 1000 and 1500 vehicles, corresponding with 10, 20 and 30 vehicles per zone over 50 zones in the studied area of NYC. To meet the demand from the extension area, 10% of vehicle fleet are initially deployed at the zone centers of the extension area, corresponding approximately to the demand from/to Suffolk County (8.9%) and Nassau County (14.6%). The parameter and the simulation setting are kept the same.

Table 13 – 14 shows that extending the service from NYC to Suffolk County or to Nassau County for the system with rideshare only would significantly increase the mean passenger journey time and mean vehicle travel time for all scenarios (+62.6% in JT and +59.1% in VTL for NYC<->Suffolk scenario with 500 vehicles, system with rideshare only, for example). The impact of extending service coverage on JT and VTL becomes less significant for the system with rideshare-transit option (with less overall JT and VTL increase). Extending the service coverage to Nassau County would be more appealing compared to the extension to Suffolk County due to its lower VTL (90.0 vs. 92.8 minutes and 61.5 vs. 65.0 minutes for fleet size 1000 and 1500 vehicles), given its higher additional customers (+1392 vs. +794). There is a slight increase (+0.34% ~ 1.31%) in mean passenger journey time when extending the service to Nassau County or Suffolk County by considering the transit-rideshare with a fleet size of 1000 or 1500 vehicles. The result concludes extending the service to Nassau County is preferred.

Table 12 Demand between OD counties during 7-9 a.m.

| O/D | Suffolk | Nassau | NYC | Total |
|---|---|---|---|---|
| Suffolk | 602 | 214 | 104 | 920 |
| Nassau | 56 | 851 | 44 | 951 |
| NYC | 88 | 497 | 8116 | 8701 |
| Total | 746 | 1562 | 8264 | 10572 |

Remark: Measured in number of individuals

Table 13 Service coverage extension analysis

| Number of vehicles | | NYC | | NYC-Suffolk | | NYC-Nassau | |
|---|---|---|---|---|---|---|---|
| | | R | R+T | R | R+T | R | R+T |
| 500 | JT | 93.3 | 52.1 | 151.7 | 67.8 | 162.3 | 69.1 |
| | VTL | 286.3 | 170.0 | 455.6 | 206.9 | 463.8 | 209.5 |
| 1000 | JT | 44.0 | 30.5 | 71.1 | 38.9 | 94.2 | 37.8 |
| | VTL | 131.9 | 76.9 | 222.2 | 92.8 | 216.2 | 90.0 |
| 1500 | JT | 35.1 | 29.7 | 52.3 | 37.8 | 48.9 | 35.6 |
| | VTL | 86.4 | 52.4 | 150.0 | 65.0 | 138.2 | 61.5 |

Remark: JT: Mean passenger journey time (in minutes), VTL: Mean vehicle travel time (in minutes).

Table 14 Mean passenger journey time for different service coverage extensions

| Number of vehicles | System | NYC | NYC-Suffolk | | NYC-Nassau | |
|---|---|---|---|---|---|---|
| | | | NYC | Suffolk | NYC | Nassau |
| 500 | R | 93.3 | 109.9 | 579.7 | 114.3 | 442.3 |
| | RT | 52.1 | 57.1 | 176.2 | 58.4 | 131.5 |
| 1000 | R | 44 | 48.3 | 305.0 | 57.3 | 309.0 |
| | RT | 30.5 | 30.9 | 121.1 | 30.7 | 79.1 |
| 1500 | R | 35.1 | 36.2 | 217.4 | 36.7 | 120.5 |
| | RT | 29.7 | 29.8 | 120.3 | 29.8 | 69.6 |

Remark: 1. R: system with rideshare only, RT: system with rideshare-transit options. 2. Measured in minutes.

## 6. Conclusions

In this study, we argue that integrating MoD fleet management (dispatch, rebalancing) with a public transport network for door-to-door service can be substantially more beneficial. This benefit is further enhanced by the careful design of anticipative algorithms to handle dispatch and relocation. Whereas first/last mile algorithms in the literature only consider linking customers with a PT station, the proposed system is flexible and considers the user's complete trip from door to door with a range of options: rideshare only, rideshare-transit-rideshare, and rideshare-transit-walk (and vice versa). This is the first study to propose integrated dynamic dispatch and idle vehicle relocation algorithms to provide door-to-door multimodal service in the presence of a PT network. A number of insights were gained from the computational experiments conducted using synthetic and realistic data instances.

- Cost savings can be substantial and benefit both users and operators, although the amount of benefit varies by type of network and demand patterns of the users. The bimodal operation can still provide significant improvements even when non-myopic algorithms do not do well in the unimodal instance. For example, the nonmyopic relocation algorithm for rideshare-only option does not make any significant improvements in the LIRR case study. On the other hand, the transit-rideshare system outperforms the rideshare-only system by 32% reduction in user journey time and 64% vehicle travel time for a 20-vehicle per zone fleet.
- While cost savings for the operator are intuitive, the savings for users are less so since they now have to wait for transit as part of their trip. The savings come from the reduction of the MoD trip lengths (28% for LIRR case study) which result in having increased fleet available to serve customers so that their average wait times for MoD service are significantly reduced. In total, there is an effective increase in the capacity of the MoD service of 4.05 when linking with the PT network for the LIRR case study.
- The users of the expanded service options are quite heterogeneous. In the LIRR case study, approximately 60% of the users would just stick with rideshare-only option, while 34% use RTW/WTR options, and 5% use RTR. The proposed algorithm provides users with all these options and allows us to identify high return opportunities. An example of this is illustrated with a hypothetical scenario of expanding to either Nassau or Suffolk County, where the algorithm is used to provide decision support. Extending the service coverage to Nassau County is preferred.

The study does have shortcomings. User costs are not quantified in terms of their perceived costs (for example, that one minute of transfer time is generally perceived to be longer than one minute of in-vehicle time, which can also differ between ride-share time and transit vehicle time) since such data is not available for a strategy that has not yet been implemented in practice. As discussed in the introduction, we chose to keep simpler assumptions to have a more straightforward comparison of algorithms on the supply side. Demand evaluation of the system therefore remains an open research question that needs to be addressed.

For future extensions, one can study an efficient algorithm to solve large-scale idle vehicle relocation problem with a grid-like zoning system. The recent work by Sayarshad and Chow (2017) showed some promising result. Another research area is the extension of the proposed strategy to consider electric or autonomous vehicles. In this setting, vehicle charging scheme and charging station constraints (availability and capacity constraints for example) need to be integrated in vehicle's routing and dispatching decisions. The customer-to-server allocation can be modified to allow different users in zone $i$ be served by different zone $j$'s, which would lead to an integral mathematical program with equilibrium constraints. Finally, integrating customer choice behavior modeling in MoD system operation policy design and revenue management could address the interactions of system performance and customer's acceptance of using the system.

**Appendix**

1. Influence of $\beta$

As noted by Hyytïa et al. (2012), the value of $\beta$ needs to be calibrated, where $\beta = 0$ suggests a myopic model. We test 31 values of $\beta$ on the performance of the system for low and high customer arrival intensities. We observed non-myopic vehicle dispatching policy with non-zero $\beta$ value can effectively improve the system performance. The effectiveness of the non-myopic vehicle dispatching policy depends on the chosen value of $\beta$. We see from Table A1 and Fig. A1 the value of $\beta$ set as $4/\bar{T}(v,x)$ and $5/\bar{T}(v,x)$ produce the most effective results in terms of mean vehicle travel time for low and high customer arrival cases.

Table A1 The influence of $\beta$ on the performance of the system

| $\beta$ | k | $\lambda = 100$ | | | $\lambda = 400$ | | |
|---|---|---|---|---|---|---|---|
| | | WT | JT | VTL | WT | JT | VTL |
| $\dfrac{k}{\bar{T}(v,x)}$ | 0 | 10.5 | 32.0 | 93.5 | 89.8 | 126.7 | 378.4 |
| | 1 | 10.9 | 32.9 | 94.0 | 86.8 | 124.1 | 365.4 |
| | 2 | 11.0 | 33.4 | 91.1 | 88.4 | 126.0 | 368.2 |
| | 3 | 10.9 | 32.6 | 93.9 | 88.5 | 125.4 | 357.1 |
| | 4 | 11.0 | 33.4 | 92.7 | **86.7** | **124.0** | **355.9** |
| | 5 | **11.6** | **34.5** | **90.6** | 92.6 | 131.9 | 371.3 |
| | 6 | 11.4 | 34.7 | 90.9 | 98.9 | 140.1 | 384.6 |
| | 7 | 11.7 | 34.7 | 91.8 | 94.2 | 133.3 | 374.9 |
| | 8 | 11.9 | 34.2 | 91.1 | 99.8 | 140.1 | 383.6 |
| | 9 | 11.9 | 34.2 | 91.1 | 93.7 | 133.5 | 367.1 |
| | 10 | 11.9 | 34.2 | 91.1 | 100.5 | 141.1 | 379.3 |
| $\dfrac{0.1k}{\bar{T}(v,x)}$ | 1 | 10.8 | 32.4 | 95.2 | 88.3 | 125.0 | 372.9 |
| | 2 | 10.5 | 31.7 | 93.7 | 88.8 | 123.9 | 370.6 |
| | 3 | 10.7 | 31.8 | 93.8 | 90.2 | 127.2 | 374.9 |
| | 4 | 10.4 | 31.8 | 93.1 | 90.8 | 127.0 | 381.2 |
| | 5 | 10.7 | 32.2 | 94.9 | 86.7 | 124.1 | 367.8 |
| | 6 | 10.6 | 32.5 | 94.2 | 89.1 | 125.7 | 373.4 |
| | 7 | 10.6 | 32.5 | 94.2 | 88.0 | 124.3 | 369.3 |
| | 8 | 10.6 | 32.5 | 94.4 | 90.0 | 125.7 | 364.8 |
| | 9 | 10.9 | 32.9 | 94.0 | 89.3 | 125.9 | 367.1 |
| | 10 | 10.9 | 32.9 | 94.0 | 85.4 | 123.7 | 362.8 |
| $\dfrac{0.01k}{\bar{T}(v,x)}$ | 1 | 10.5 | 32.0 | 93.5 | 87.9 | 125.1 | 374.2 |
| | 2 | 10.5 | 32.0 | 93.5 | 87.9 | 125.1 | 374.2 |
| | 3 | 10.8 | 32.2 | 94.5 | 87.9 | 125.1 | 374.2 |
| | 4 | 10.8 | 32.3 | 94.5 | 87.9 | 125.1 | 374.2 |
| | 5 | 10.8 | 32.4 | 95.2 | 90.4 | 126.4 | 376.6 |
| | 6 | 10.8 | 32.4 | 95.2 | 90.0 | 125.8 | 375.1 |
| | 7 | 10.8 | 32.4 | 95.2 | 90.0 | 125.8 | 375.1 |
| | 8 | 10.8 | 32.4 | 95.2 | 90.0 | 125.8 | 375.1 |
| | 9 | 10.8 | 32.4 | 95.2 | 90.0 | 125.8 | 375.1 |
| | 10 | 10.8 | 32.4 | 95.2 | 89.7 | 124.6 | 370.0 |

Remark: 1. WT: Mean passenger waiting time, JT: Mean passenger journey time (waiting and riding time), VTL: Mean vehicle travel time. 2. $\bar{T}(v,x)$ is mean vehicle travel time when using $\beta = 0$. 3. Measured in minutes.

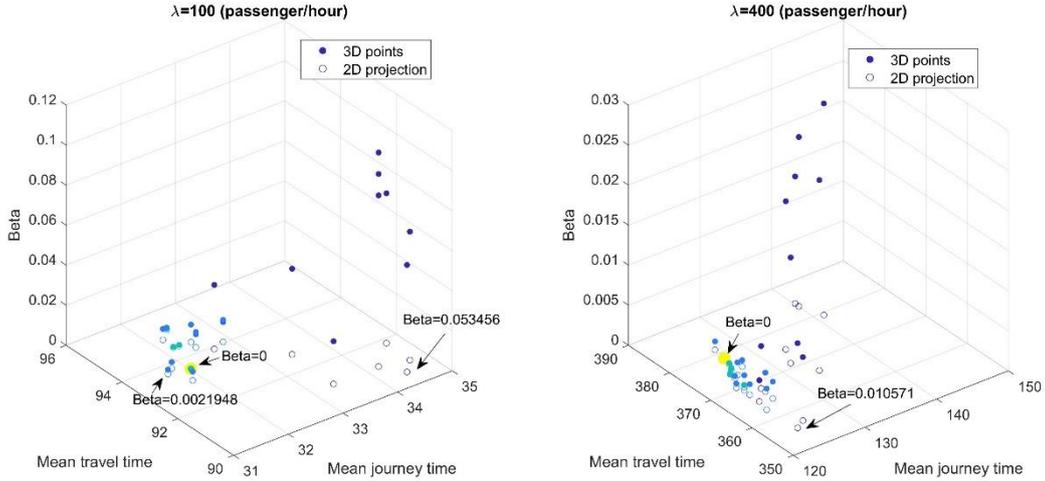

Fig. A1. The influence of $\beta$ on the performance of non-myopic vehicle dispatching policy (Left: $\lambda = 100$ passengers/hour; Right: $\lambda = 400$ passengers/hour).

2. Influence of $\theta$

We vary the value of $\theta$ from 0 to 4 to assess its influence on the performance of the system. The result is shown in Fig. A2. For low customer arrival intensity case, $\theta$ has little impact on the system performance. However, for high customer arrival rate, $\theta$ influences the effectiveness of the idle vehicle relocation. Using $\theta \geq 1.2$ can reduce mean passenger journey time (-7.1%) and mean operation cost (-2.6%) compared to the benchmark (no idle vehicle relocation). The result suggests that the value of $\theta$ needs to be calibrated to make vehicle rebalancing effective.

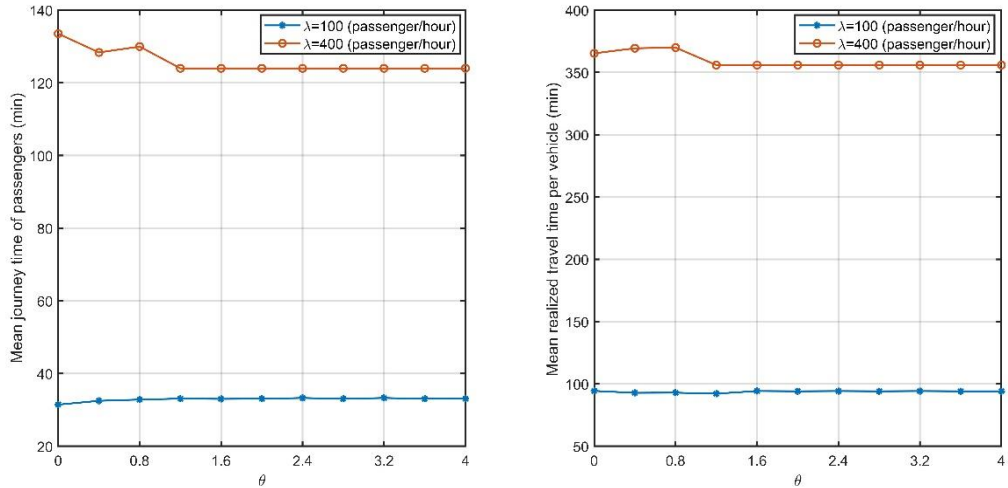

Fig. A2. Influence of $\theta$ on the performance of the system (Left: mean journey time of passengers; Right: mean realized travel time per vehicle)

3. Impact of different idle vehicle relocation policy

We compare the proposed idle vehicle relocation policy P2 with three other alternatives. To show how customer arrival rate, service rate and idle vehicle availability (number of servers) influence the performance of relocation policy, we set up four experiments with different customer arrival rates, ranging from 50 to 400 customers/hour. The results are shown in Table A2 and Fig. A3. When travel demand is not too high (i.e. $\lambda = 50$ and $\lambda = 100$) with idle vehicles available for rebalancing, the non-myopic relocation policy performs better than the other relocation policies (upper part of Table A2). However, when customer arrival rate is too high (i.e. $\lambda = 200$ and $\lambda = 400$), all vehicles become busy after 20 minutes ($\lambda = 200$) and 40 minutes ($\lambda = 400$) (see lower part of Fig. A3), resulting in similar system performance when different relocation policies are applied (see lower part of Table A2).

By using the M/M/s approximation and assuming service times are exponentially distributed, the non-myopic relocation can obtain non-dominated solutions compared to the myopic relocation in terms of the integrated performance measures of passenger wait time, journey time, and vehicle trip lengths.

Table A2 Influence of different idle vehicle relocation policies

| Policy | $\lambda = 50$ | | | | $\lambda = 100$ | | | |
|---|---|---|---|---|---|---|---|---|
| | WT | JT | VTL | Comp. time (min.) | WT | JT | VTL | Comp. time (min.) |
| Waiting policy | 13.0 | 36.5 | 56.4 | 1.4 | 11.6 | 34.5 | 90.6 | 2.4 |
| Busiest zone policy | 13.0 | 36.5 | 56.4 | 1.0 | 11.6 | 34.5 | 90.6 | 2.0 |
| Myopic relocation | 10.9 | 34.3 | 55.1 | 1.2 | 11.9 | 35.6 | 91.3 | 2.3 |
| Non-myopic relocation | **11.1** | **34.5** | **54.9** | 2.7 | 10.2 | 33.1 | 92.3 | 2.4 |
| | $\lambda = 200$ | | | | $\lambda = 400$ | | | |
| Waiting policy | 25.5 | 52.4 | 174.2 | 4.5 | 86.7 | 124.0 | 355.9 | 10.5 |
| Busiest zone policy | 25.5 | 52.4 | 174.2 | 4.3 | 86.7 | 124.0 | 355.9 | 10.9 |
| Myopic relocation | 24.9 | 52.8 | 175.4 | 4.3 | 86.7 | 124.0 | 355.9 | 11.5 |
| Non-myopic relocation | **24.9** | **52.8** | **175.4** | 4.4 | 86.7 | 124.0 | 355.9 | 11.4 |

Remark: 1. WT: Mean passenger waiting time, JT: Mean passenger journey time (waiting and riding time), VTL: Mean vehicle travel time. 2. Idle vehicles in transition to its assigned zone are allowed to pick up new customers. 2. Computational time is the total simulation time of the case study. 3. The average computational time for solving a single P2 instance for myopic relocation policy is 1 second ( both $\lambda = 50$ and $\lambda = 400$). For non-myopic relocation policy, it is 3.2 sec. ($\lambda = 50$) and 2.7 sec. ($\lambda = 400$), respectively.

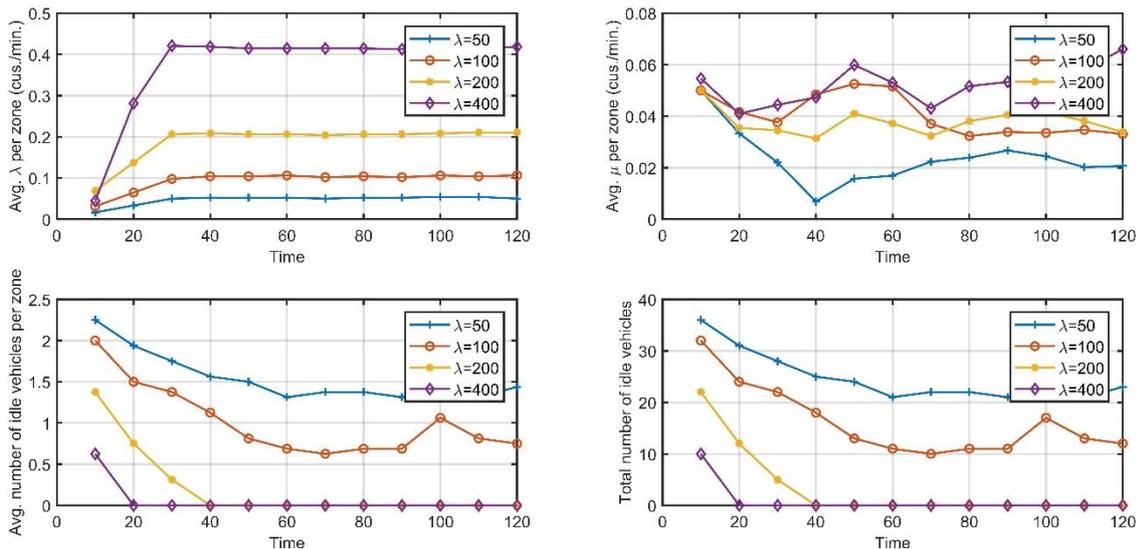

Fig. A3. Evolution of $\lambda$, $\mu$ and average number of idle vehicles per zone over time for the system using non-myopic relocation policy.

4. Influence of idle vehicle en-route switching policy

We further compare the performance of en-route switching policy for idle vehicles in transition to its relocated zones. Two policies are compared: the first one allows in-transition idle vehicles to pick up new customers. The second one doesn't allow en-route switching behavior when repositioning. We test on the non-myopic relocation model with increasing arrival intensity, the result shows allowing en-route switching to pick up new customers when idle vehicles in transition can reduce system operation cost.

Table A3 Influence of idle vehicle en-route switching policy

| Idle vehicles en-route switching to pick up new customers? | $\lambda = 50$ | | | $\lambda = 100$ | | |
|---|---|---|---|---|---|---|
| | Mean passenger waiting time | Mean passenger journey time | Mean vehicle travel time | Mean passenger waiting time | Mean passenger journey time | Mean vehicle travel time |
| Not allowed | 13.3 | 37.4 | 57.5 | 11.6 | 34.5 | 90.6 |
| Allowed | **11.1** | **34.5** | **54.9** | **10.2** | **33.1** | **92.3** |
| | $\lambda = 200$ | | | $\lambda = 400$ | | |
| Not allowed | 28.1 | 56.4 | 178.8 | 86.7 | 124.0 | 355.9 |
| Allowed | **24.9** | **52.8** | **175.4** | **86.7** | **124.0** | **355.9** |

Remark: Measured in minutes.

5. Influence of adaptive learning of service rate $\mu$ and dynamic zone centroid adjustment policy

We compare the performance with/without (a) adaptive learning of service rate $\mu$ and (b) dynamic zone centroid adjustment policy under different customer arrival intensities. The result shows adopting these strategies could effectively improve the system performance (Table A4). We found dynamic zone centroid adjustment policy have more significant benefice compared to adaptive learning of service rate $\mu$.

Table A4 Influence of service rate ($\mu$) learning and dynamically zone centroids adjustment

| Dynamic service rate learning and zone centroids updates | $\lambda = 50$ | | | $\lambda = 100$ | | |
|---|---|---|---|---|---|---|
| | Mean passenger waiting time | Mean passenger journey time | Mean vehicle travel time | Mean passenger waiting time | Mean passenger journey time | Mean vehicle travel time |
| No (a) | 11.7 (+5.4%) | 35.3 (+2.3%) | 56.5 (+2.9%) | 11.2 (+9.8%) | 33.9 (+2.4%) | 94.1 (+2.0%) |
| No (b) | 12.5 (+12.6%) | 36.6 (+6.1%) | 57.3 (+4.4%) | 11 (+7.8%) | 35.3 (+6.6%) | 92.4 (+0.1%) |
| With (a) and (b) | 11.1 | 34.5 | 54.9 | 10.2 | 33.1 | 92.3 |
| | $\lambda = 200$ | | | $\lambda = 400$ | | |
| No (a) | 24.9 (+0%) | 52.8 (+0%) | 175.4 (+0%) | 86.7 (+0%) | 124 (+0%) | 355.9 (+0%) |
| No (b) | 26.7 (+7.2%) | 54.2 (+2.7%) | 177.6 (+1.3) | 86.7 (+0%) | 124 (+0%) | 355.9 (+0%) |
| With (a) and (b) | **24.9** | **52.8** | **175.4** | **86.7** | **124.0** | **355.9** |

Remark: (a): adaptive $\mu$ learning. (b): zone centroids adjustment. Measured in minutes.